\documentclass[10pt,conference, anonymous,review]{IEEEtran}
\usepackage{xspace}
\usepackage{hyperref}
\usepackage{multirow}
\usepackage{enumitem}
\usepackage{float}
\usepackage{color, colortbl}
\usepackage{booktabs} 
\usepackage{pgfkeys,pgfmath}
\usepackage{graphicx}
\usepackage{caption}
\usepackage{subcaption}
\usepackage{booktabs,colortbl}
\usepackage{varwidth}
\usepackage{amsmath}
\usepackage{textcomp}
\usepackage{siunitx}
\usepackage{balance}
\usepackage[utf8]{inputenc}
\usepackage[T1]{fontenc}
\usepackage{microtype}

\newcommand{\printpercent}[3][2]{%
    \pgfmathdivide{#2}{#3}%
    \pgfmathmultiply{\pgfmathresult}{100}%
    \SI[round-mode=places,round-precision=#1]{\pgfmathresult}{\percent}
}%

\newcommand{\udensdot}[1]{%
    \tikz[baseline=(todotted.base)]{
        \node[inner sep=1pt,outer sep=0pt] (todotted) {#1};
        \draw[densely dotted] (todotted.south west) -- (todotted.south east);
    }%
}%

\newcommand{\post}[1]{\href{http://stackoverflow.com/questions/#1}{\udensdot{#1}}}

\newcommand{\answer}[1]{\href{http://stackoverflow.com/a/#1}{\udensdot{#1}}}

\newcommand{\wordpattern}{\texttt{wordpattern}\xspace}
\newcommand{\simpleif}{\texttt{simpleif}\xspace}
\newcommand{\condinsight}{\texttt{contextif}\xspace}
\newcommand{\lexrank}{\texttt{lexrank}\xspace}
\newcommand{\Figref}[1]{Figure~\ref{#1}} 

\newcommand{\navcue}{navigational cue\xspace}

\newcommand{\subscript}[2]{$#1 _ {#2}$}

\usepackage[absolute,overlay]{textpos}
\usepackage{tikz}
\usepackage{xcolor}
\usepackage{calc}

\newcounter{findingctr}

\newcommand{\finding}[2]{\refstepcounter{findingctr}\emph{Answer to #1:\label{box:#2}}}

\newlength{\boxw}
\newlength{\boxh}
\newlength{\shadowsize}
\newlength{\boxroundness}
\newlength{\tmpa}
\newsavebox{\shadowblockbox}

\newenvironment{findingenv}[2]%
{\vspace{0.2cm}\noindent
\begin{lrbox}{
\shadowblockbox
}
\begin{minipage}{.98\columnwidth}
\finding{#1}{#2}~}%
{\end{minipage}\end{lrbox}%
\settowidth{\boxw}{\usebox{\shadowblockbox}}%
\settodepth{\tmpa}{\usebox{\shadowblockbox}}%
\settoheight{\boxh}{\usebox{\shadowblockbox}}%
\addtolength{\boxh}{\tmpa}%
\begin{tikzpicture}
\addtolength{\boxw}{\boxroundness * 2}
\addtolength{\boxh}{\boxroundness * 2}

\foreach \x in {0,.05,...,1}
{
\setlength{\tmpa}{\shadowsize * \real{\x}}
\fill[xshift=\shadowsize - 1pt,yshift=-\shadowsize + 
1pt,black,opacity=.04,rounded corners=\boxroundness] 
(\tmpa, \tmpa) rectangle +(\boxw - \tmpa - \tmpa, \boxh - \tmpa - 
\tmpa);
}

\filldraw[fill=white!50, draw=black!80, rounded corners=\boxroundness] (0, 
0) rectangle (\boxw, \boxh);
\draw node[xshift=\boxroundness,yshift=\boxroundness,inner sep=0pt,outer 
sep=0pt,anchor=south west] (0,0) {\usebox{\shadowblockbox}};
\end{tikzpicture}\vspace{0cm}%
}

\pgfkeyssetvalue{original_participants}{59}
\pgfkeyssetvalue{filtered_out_participants}{16}
\pgfkeyssetvalue{final_num_participants}{43}
\pgfkeyssetvalue{total_threads_eval}{20}
\pgfkeyssetvalue{total_sentence_eval}{76}
\pgfkeyssetvalue{number_simpleif_sentences}{49}
\pgfkeyssetvalue{number_condinsight_sentences}{20}
\pgfkeyssetvalue{number_wordpattern_sentences}{21}
\pgfkeyssetvalue{number_lexrank_sentences}{20}
\pgfkeyssetvalue{rating_per_thread_min}{3}
\pgfkeyssetvalue{rating_per_thread_max}{10}
\pgfkeyssetvalue{rating_per_thread_median}{7}
\pgfkeyssetvalue{rating_per_thread_mode}{7}
\pgfkeyssetvalue{ground_truth_size_q8_limit_1}{76}
\pgfkeyssetvalue{neg_ground_truth_size_q8_limit_1}{0}
\pgfkeyssetvalue{total_pos_tech1_q8_threshold_1}{49}
\pgfkeyssetvalue{total_neg_tech1_q8_threshold_0}{0}
\pgfkeyssetvalue{mean_sentence_rating_tech1_q8}{3}
\pgfkeyssetvalue{precision_tech1_q8_threshold_1}{1.00}
\pgfkeyssetvalue{recall_tech1_q8_threshold_1}{0.64}
\pgfkeyssetvalue{total_pos_tech2_q8_threshold_1}{20}
\pgfkeyssetvalue{total_neg_tech2_q8_threshold_0}{0}
\pgfkeyssetvalue{mean_sentence_rating_tech2_q8}{3}
\pgfkeyssetvalue{precision_tech2_q8_threshold_1}{1.00}
\pgfkeyssetvalue{recall_tech2_q8_threshold_1}{0.26}
\pgfkeyssetvalue{total_pos_tech3_q8_threshold_1}{21}
\pgfkeyssetvalue{total_neg_tech3_q8_threshold_0}{0}
\pgfkeyssetvalue{mean_sentence_rating_tech3_q8}{3}
\pgfkeyssetvalue{precision_tech3_q8_threshold_1}{1.00}
\pgfkeyssetvalue{recall_tech3_q8_threshold_1}{0.28}
\pgfkeyssetvalue{total_pos_tech6_q8_threshold_1}{20}
\pgfkeyssetvalue{total_neg_tech6_q8_threshold_0}{0}
\pgfkeyssetvalue{mean_sentence_rating_tech6_q8}{3}
\pgfkeyssetvalue{precision_tech6_q8_threshold_1}{1.00}
\pgfkeyssetvalue{recall_tech6_q8_threshold_1}{0.26}
\pgfkeyssetvalue{ground_truth_size_q8_limit_2}{76}
\pgfkeyssetvalue{neg_ground_truth_size_q8_limit_2}{0}
\pgfkeyssetvalue{total_pos_tech1_q8_threshold_2}{49}
\pgfkeyssetvalue{total_neg_tech1_q8_threshold_1}{0}
\pgfkeyssetvalue{mean_sentence_rating_tech1_q8}{3}
\pgfkeyssetvalue{precision_tech1_q8_threshold_2}{1.00}
\pgfkeyssetvalue{recall_tech1_q8_threshold_2}{0.64}
\pgfkeyssetvalue{total_pos_tech2_q8_threshold_2}{20}
\pgfkeyssetvalue{total_neg_tech2_q8_threshold_1}{0}
\pgfkeyssetvalue{mean_sentence_rating_tech2_q8}{3}
\pgfkeyssetvalue{precision_tech2_q8_threshold_2}{1.00}
\pgfkeyssetvalue{recall_tech2_q8_threshold_2}{0.26}
\pgfkeyssetvalue{total_pos_tech3_q8_threshold_2}{21}
\pgfkeyssetvalue{total_neg_tech3_q8_threshold_1}{0}
\pgfkeyssetvalue{mean_sentence_rating_tech3_q8}{3}
\pgfkeyssetvalue{precision_tech3_q8_threshold_2}{1.00}
\pgfkeyssetvalue{recall_tech3_q8_threshold_2}{0.28}
\pgfkeyssetvalue{total_pos_tech6_q8_threshold_2}{20}
\pgfkeyssetvalue{total_neg_tech6_q8_threshold_1}{0}
\pgfkeyssetvalue{mean_sentence_rating_tech6_q8}{3}
\pgfkeyssetvalue{precision_tech6_q8_threshold_2}{1.00}
\pgfkeyssetvalue{recall_tech6_q8_threshold_2}{0.26}
\pgfkeyssetvalue{ground_truth_size_q8_limit_3}{51}
\pgfkeyssetvalue{neg_ground_truth_size_q8_limit_3}{21}
\pgfkeyssetvalue{total_pos_tech1_q8_threshold_3}{32}
\pgfkeyssetvalue{total_neg_tech1_q8_threshold_2}{14}
\pgfkeyssetvalue{mean_sentence_rating_tech1_q8}{3}
\pgfkeyssetvalue{precision_tech1_q8_threshold_3}{0.65}
\pgfkeyssetvalue{recall_tech1_q8_threshold_3}{0.63}
\pgfkeyssetvalue{total_pos_tech2_q8_threshold_3}{13}
\pgfkeyssetvalue{total_neg_tech2_q8_threshold_2}{6}
\pgfkeyssetvalue{mean_sentence_rating_tech2_q8}{3}
\pgfkeyssetvalue{precision_tech2_q8_threshold_3}{0.65}
\pgfkeyssetvalue{recall_tech2_q8_threshold_3}{0.25}
\pgfkeyssetvalue{total_pos_tech3_q8_threshold_3}{16}
\pgfkeyssetvalue{total_neg_tech3_q8_threshold_2}{4}
\pgfkeyssetvalue{mean_sentence_rating_tech3_q8}{3}
\pgfkeyssetvalue{precision_tech3_q8_threshold_3}{0.76}
\pgfkeyssetvalue{recall_tech3_q8_threshold_3}{0.31}
\pgfkeyssetvalue{total_pos_tech6_q8_threshold_3}{16}
\pgfkeyssetvalue{total_neg_tech6_q8_threshold_2}{3}
\pgfkeyssetvalue{mean_sentence_rating_tech6_q8}{3}
\pgfkeyssetvalue{precision_tech6_q8_threshold_3}{0.80}
\pgfkeyssetvalue{recall_tech6_q8_threshold_3}{0.31}
\pgfkeyssetvalue{ground_truth_size_q9_limit_1}{76}
\pgfkeyssetvalue{neg_ground_truth_size_q9_limit_1}{0}
\pgfkeyssetvalue{total_pos_tech1_q9_threshold_1}{49}
\pgfkeyssetvalue{total_neg_tech1_q9_threshold_0}{0}
\pgfkeyssetvalue{mean_sentence_rating_tech1_q9}{4}
\pgfkeyssetvalue{precision_tech1_q9_threshold_1}{1.00}
\pgfkeyssetvalue{recall_tech1_q9_threshold_1}{0.64}
\pgfkeyssetvalue{total_pos_tech2_q9_threshold_1}{20}
\pgfkeyssetvalue{total_neg_tech2_q9_threshold_0}{0}
\pgfkeyssetvalue{mean_sentence_rating_tech2_q9}{4}
\pgfkeyssetvalue{precision_tech2_q9_threshold_1}{1.00}
\pgfkeyssetvalue{recall_tech2_q9_threshold_1}{0.26}
\pgfkeyssetvalue{total_pos_tech3_q9_threshold_1}{21}
\pgfkeyssetvalue{total_neg_tech3_q9_threshold_0}{0}
\pgfkeyssetvalue{mean_sentence_rating_tech3_q9}{4}
\pgfkeyssetvalue{precision_tech3_q9_threshold_1}{1.00}
\pgfkeyssetvalue{recall_tech3_q9_threshold_1}{0.28}
\pgfkeyssetvalue{total_pos_tech6_q9_threshold_1}{20}
\pgfkeyssetvalue{total_neg_tech6_q9_threshold_0}{0}
\pgfkeyssetvalue{mean_sentence_rating_tech6_q9}{4}
\pgfkeyssetvalue{precision_tech6_q9_threshold_1}{1.00}
\pgfkeyssetvalue{recall_tech6_q9_threshold_1}{0.26}
\pgfkeyssetvalue{ground_truth_size_q9_limit_2}{74}
\pgfkeyssetvalue{neg_ground_truth_size_q9_limit_2}{1}
\pgfkeyssetvalue{total_pos_tech1_q9_threshold_2}{47}
\pgfkeyssetvalue{total_neg_tech1_q9_threshold_1}{1}
\pgfkeyssetvalue{mean_sentence_rating_tech1_q9}{4}
\pgfkeyssetvalue{precision_tech1_q9_threshold_2}{0.96}
\pgfkeyssetvalue{recall_tech1_q9_threshold_2}{0.64}
\pgfkeyssetvalue{total_pos_tech2_q9_threshold_2}{19}
\pgfkeyssetvalue{total_neg_tech2_q9_threshold_1}{1}
\pgfkeyssetvalue{mean_sentence_rating_tech2_q9}{4}
\pgfkeyssetvalue{precision_tech2_q9_threshold_2}{0.95}
\pgfkeyssetvalue{recall_tech2_q9_threshold_2}{0.26}
\pgfkeyssetvalue{total_pos_tech3_q9_threshold_2}{21}
\pgfkeyssetvalue{total_neg_tech3_q9_threshold_1}{0}
\pgfkeyssetvalue{mean_sentence_rating_tech3_q9}{4}
\pgfkeyssetvalue{precision_tech3_q9_threshold_2}{1.00}
\pgfkeyssetvalue{recall_tech3_q9_threshold_2}{0.28}
\pgfkeyssetvalue{total_pos_tech6_q9_threshold_2}{20}
\pgfkeyssetvalue{total_neg_tech6_q9_threshold_1}{0}
\pgfkeyssetvalue{mean_sentence_rating_tech6_q9}{4}
\pgfkeyssetvalue{precision_tech6_q9_threshold_2}{1.00}
\pgfkeyssetvalue{recall_tech6_q9_threshold_2}{0.27}
\pgfkeyssetvalue{ground_truth_size_q9_limit_3}{68}
\pgfkeyssetvalue{neg_ground_truth_size_q9_limit_3}{7}
\pgfkeyssetvalue{total_pos_tech1_q9_threshold_3}{44}
\pgfkeyssetvalue{total_neg_tech1_q9_threshold_2}{5}
\pgfkeyssetvalue{mean_sentence_rating_tech1_q9}{4}
\pgfkeyssetvalue{precision_tech1_q9_threshold_3}{0.90}
\pgfkeyssetvalue{recall_tech1_q9_threshold_3}{0.65}
\pgfkeyssetvalue{total_pos_tech2_q9_threshold_3}{17}
\pgfkeyssetvalue{total_neg_tech2_q9_threshold_2}{3}
\pgfkeyssetvalue{mean_sentence_rating_tech2_q9}{4}
\pgfkeyssetvalue{precision_tech2_q9_threshold_3}{0.85}
\pgfkeyssetvalue{recall_tech2_q9_threshold_3}{0.25}
\pgfkeyssetvalue{total_pos_tech3_q9_threshold_3}{19}
\pgfkeyssetvalue{total_neg_tech3_q9_threshold_2}{1}
\pgfkeyssetvalue{mean_sentence_rating_tech3_q9}{4}
\pgfkeyssetvalue{precision_tech3_q9_threshold_3}{0.90}
\pgfkeyssetvalue{recall_tech3_q9_threshold_3}{0.28}
\pgfkeyssetvalue{total_pos_tech6_q9_threshold_3}{19}
\pgfkeyssetvalue{total_neg_tech6_q9_threshold_2}{1}
\pgfkeyssetvalue{mean_sentence_rating_tech6_q9}{4}
\pgfkeyssetvalue{precision_tech6_q9_threshold_3}{0.95}
\pgfkeyssetvalue{recall_tech6_q9_threshold_3}{0.28}
\pgfkeyssetvalue{ground_truth_size_q9_limit_4}{49}
\pgfkeyssetvalue{neg_ground_truth_size_q9_limit_4}{23}
\pgfkeyssetvalue{total_pos_tech1_q9_threshold_4}{30}
\pgfkeyssetvalue{total_neg_tech1_q9_threshold_3}{16}
\pgfkeyssetvalue{mean_sentence_rating_tech1_q9}{4}
\pgfkeyssetvalue{precision_tech1_q9_threshold_4}{0.61}
\pgfkeyssetvalue{recall_tech1_q9_threshold_4}{0.61}
\pgfkeyssetvalue{total_pos_tech2_q9_threshold_4}{9}
\pgfkeyssetvalue{total_neg_tech2_q9_threshold_3}{9}
\pgfkeyssetvalue{mean_sentence_rating_tech2_q9}{4}
\pgfkeyssetvalue{precision_tech2_q9_threshold_4}{0.45}
\pgfkeyssetvalue{recall_tech2_q9_threshold_4}{0.18}
\pgfkeyssetvalue{total_pos_tech3_q9_threshold_4}{17}
\pgfkeyssetvalue{total_neg_tech3_q9_threshold_3}{4}
\pgfkeyssetvalue{mean_sentence_rating_tech3_q9}{4}
\pgfkeyssetvalue{precision_tech3_q9_threshold_4}{0.81}
\pgfkeyssetvalue{recall_tech3_q9_threshold_4}{0.35}
\pgfkeyssetvalue{total_pos_tech6_q9_threshold_4}{14}
\pgfkeyssetvalue{total_neg_tech6_q9_threshold_3}{4}
\pgfkeyssetvalue{mean_sentence_rating_tech6_q9}{4}
\pgfkeyssetvalue{precision_tech6_q9_threshold_4}{0.70}
\pgfkeyssetvalue{recall_tech6_q9_threshold_4}{0.29}
\pgfkeyssetvalue{ground_truth_size_q9_limit_5}{6}
\pgfkeyssetvalue{neg_ground_truth_size_q9_limit_5}{64}
\pgfkeyssetvalue{total_pos_tech1_q9_threshold_5}{6}
\pgfkeyssetvalue{total_neg_tech1_q9_threshold_4}{38}
\pgfkeyssetvalue{mean_sentence_rating_tech1_q9}{4}
\pgfkeyssetvalue{precision_tech1_q9_threshold_5}{0.12}
\pgfkeyssetvalue{recall_tech1_q9_threshold_5}{1.00}
\pgfkeyssetvalue{total_pos_tech2_q9_threshold_5}{3}
\pgfkeyssetvalue{total_neg_tech2_q9_threshold_4}{15}
\pgfkeyssetvalue{mean_sentence_rating_tech2_q9}{4}
\pgfkeyssetvalue{precision_tech2_q9_threshold_5}{0.15}
\pgfkeyssetvalue{recall_tech2_q9_threshold_5}{0.50}
\pgfkeyssetvalue{total_pos_tech3_q9_threshold_5}{1}
\pgfkeyssetvalue{total_neg_tech3_q9_threshold_4}{18}
\pgfkeyssetvalue{mean_sentence_rating_tech3_q9}{4}
\pgfkeyssetvalue{precision_tech3_q9_threshold_5}{0.05}
\pgfkeyssetvalue{recall_tech3_q9_threshold_5}{0.17}
\pgfkeyssetvalue{total_pos_tech6_q9_threshold_5}{1}
\pgfkeyssetvalue{total_neg_tech6_q9_threshold_4}{16}
\pgfkeyssetvalue{mean_sentence_rating_tech6_q9}{4}
\pgfkeyssetvalue{precision_tech6_q9_threshold_5}{0.05}
\pgfkeyssetvalue{recall_tech6_q9_threshold_5}{0.17}
\pgfkeyssetvalue{ground_truth_size_q10_limit_1}{76}
\pgfkeyssetvalue{neg_ground_truth_size_q10_limit_1}{0}
\pgfkeyssetvalue{total_pos_tech1_q10_threshold_1}{49}
\pgfkeyssetvalue{total_neg_tech1_q10_threshold_0}{0}
\pgfkeyssetvalue{mean_sentence_rating_tech1_q10}{4}
\pgfkeyssetvalue{precision_tech1_q10_threshold_1}{1.00}
\pgfkeyssetvalue{recall_tech1_q10_threshold_1}{0.64}
\pgfkeyssetvalue{total_pos_tech2_q10_threshold_1}{20}
\pgfkeyssetvalue{total_neg_tech2_q10_threshold_0}{0}
\pgfkeyssetvalue{mean_sentence_rating_tech2_q10}{4}
\pgfkeyssetvalue{precision_tech2_q10_threshold_1}{1.00}
\pgfkeyssetvalue{recall_tech2_q10_threshold_1}{0.26}
\pgfkeyssetvalue{total_pos_tech3_q10_threshold_1}{21}
\pgfkeyssetvalue{total_neg_tech3_q10_threshold_0}{0}
\pgfkeyssetvalue{mean_sentence_rating_tech3_q10}{4}
\pgfkeyssetvalue{precision_tech3_q10_threshold_1}{1.00}
\pgfkeyssetvalue{recall_tech3_q10_threshold_1}{0.28}
\pgfkeyssetvalue{total_pos_tech6_q10_threshold_1}{20}
\pgfkeyssetvalue{total_neg_tech6_q10_threshold_0}{0}
\pgfkeyssetvalue{mean_sentence_rating_tech6_q10}{4}
\pgfkeyssetvalue{precision_tech6_q10_threshold_1}{1.00}
\pgfkeyssetvalue{recall_tech6_q10_threshold_1}{0.26}
\pgfkeyssetvalue{ground_truth_size_q10_limit_2}{75}
\pgfkeyssetvalue{neg_ground_truth_size_q10_limit_2}{1}
\pgfkeyssetvalue{total_pos_tech1_q10_threshold_2}{48}
\pgfkeyssetvalue{total_neg_tech1_q10_threshold_1}{1}
\pgfkeyssetvalue{mean_sentence_rating_tech1_q10}{4}
\pgfkeyssetvalue{precision_tech1_q10_threshold_2}{0.98}
\pgfkeyssetvalue{recall_tech1_q10_threshold_2}{0.64}
\pgfkeyssetvalue{total_pos_tech2_q10_threshold_2}{19}
\pgfkeyssetvalue{total_neg_tech2_q10_threshold_1}{1}
\pgfkeyssetvalue{mean_sentence_rating_tech2_q10}{4}
\pgfkeyssetvalue{precision_tech2_q10_threshold_2}{0.95}
\pgfkeyssetvalue{recall_tech2_q10_threshold_2}{0.25}
\pgfkeyssetvalue{total_pos_tech3_q10_threshold_2}{21}
\pgfkeyssetvalue{total_neg_tech3_q10_threshold_1}{0}
\pgfkeyssetvalue{mean_sentence_rating_tech3_q10}{4}
\pgfkeyssetvalue{precision_tech3_q10_threshold_2}{1.00}
\pgfkeyssetvalue{recall_tech3_q10_threshold_2}{0.28}
\pgfkeyssetvalue{total_pos_tech6_q10_threshold_2}{20}
\pgfkeyssetvalue{total_neg_tech6_q10_threshold_1}{0}
\pgfkeyssetvalue{mean_sentence_rating_tech6_q10}{4}
\pgfkeyssetvalue{precision_tech6_q10_threshold_2}{1.00}
\pgfkeyssetvalue{recall_tech6_q10_threshold_2}{0.27}
\pgfkeyssetvalue{ground_truth_size_q10_limit_3}{68}
\pgfkeyssetvalue{neg_ground_truth_size_q10_limit_3}{6}
\pgfkeyssetvalue{total_pos_tech1_q10_threshold_3}{43}
\pgfkeyssetvalue{total_neg_tech1_q10_threshold_2}{5}
\pgfkeyssetvalue{mean_sentence_rating_tech1_q10}{4}
\pgfkeyssetvalue{precision_tech1_q10_threshold_3}{0.88}
\pgfkeyssetvalue{recall_tech1_q10_threshold_3}{0.63}
\pgfkeyssetvalue{total_pos_tech2_q10_threshold_3}{17}
\pgfkeyssetvalue{total_neg_tech2_q10_threshold_2}{3}
\pgfkeyssetvalue{mean_sentence_rating_tech2_q10}{4}
\pgfkeyssetvalue{precision_tech2_q10_threshold_3}{0.85}
\pgfkeyssetvalue{recall_tech2_q10_threshold_3}{0.25}
\pgfkeyssetvalue{total_pos_tech3_q10_threshold_3}{19}
\pgfkeyssetvalue{total_neg_tech3_q10_threshold_2}{1}
\pgfkeyssetvalue{mean_sentence_rating_tech3_q10}{4}
\pgfkeyssetvalue{precision_tech3_q10_threshold_3}{0.90}
\pgfkeyssetvalue{recall_tech3_q10_threshold_3}{0.28}
\pgfkeyssetvalue{total_pos_tech6_q10_threshold_3}{20}
\pgfkeyssetvalue{total_neg_tech6_q10_threshold_2}{0}
\pgfkeyssetvalue{mean_sentence_rating_tech6_q10}{4}
\pgfkeyssetvalue{precision_tech6_q10_threshold_3}{1.00}
\pgfkeyssetvalue{recall_tech6_q10_threshold_3}{0.29}
\pgfkeyssetvalue{ground_truth_size_q10_limit_4}{48}
\pgfkeyssetvalue{neg_ground_truth_size_q10_limit_4}{22}
\pgfkeyssetvalue{total_pos_tech1_q10_threshold_4}{29}
\pgfkeyssetvalue{total_neg_tech1_q10_threshold_3}{15}
\pgfkeyssetvalue{mean_sentence_rating_tech1_q10}{4}
\pgfkeyssetvalue{precision_tech1_q10_threshold_4}{0.59}
\pgfkeyssetvalue{recall_tech1_q10_threshold_4}{0.60}
\pgfkeyssetvalue{total_pos_tech2_q10_threshold_4}{11}
\pgfkeyssetvalue{total_neg_tech2_q10_threshold_3}{7}
\pgfkeyssetvalue{mean_sentence_rating_tech2_q10}{4}
\pgfkeyssetvalue{precision_tech2_q10_threshold_4}{0.55}
\pgfkeyssetvalue{recall_tech2_q10_threshold_4}{0.23}
\pgfkeyssetvalue{total_pos_tech3_q10_threshold_4}{16}
\pgfkeyssetvalue{total_neg_tech3_q10_threshold_3}{5}
\pgfkeyssetvalue{mean_sentence_rating_tech3_q10}{4}
\pgfkeyssetvalue{precision_tech3_q10_threshold_4}{0.76}
\pgfkeyssetvalue{recall_tech3_q10_threshold_4}{0.33}
\pgfkeyssetvalue{total_pos_tech6_q10_threshold_4}{16}
\pgfkeyssetvalue{total_neg_tech6_q10_threshold_3}{2}
\pgfkeyssetvalue{mean_sentence_rating_tech6_q10}{4}
\pgfkeyssetvalue{precision_tech6_q10_threshold_4}{0.80}
\pgfkeyssetvalue{recall_tech6_q10_threshold_4}{0.33}
\pgfkeyssetvalue{ground_truth_size_q10_limit_5}{4}
\pgfkeyssetvalue{neg_ground_truth_size_q10_limit_5}{64}
\pgfkeyssetvalue{total_pos_tech1_q10_threshold_5}{4}
\pgfkeyssetvalue{total_neg_tech1_q10_threshold_4}{37}
\pgfkeyssetvalue{mean_sentence_rating_tech1_q10}{4}
\pgfkeyssetvalue{precision_tech1_q10_threshold_5}{0.08}
\pgfkeyssetvalue{recall_tech1_q10_threshold_5}{1.00}
\pgfkeyssetvalue{total_pos_tech2_q10_threshold_5}{1}
\pgfkeyssetvalue{total_neg_tech2_q10_threshold_4}{16}
\pgfkeyssetvalue{mean_sentence_rating_tech2_q10}{4}
\pgfkeyssetvalue{precision_tech2_q10_threshold_5}{0.05}
\pgfkeyssetvalue{recall_tech2_q10_threshold_5}{0.25}
\pgfkeyssetvalue{total_pos_tech3_q10_threshold_5}{1}
\pgfkeyssetvalue{total_neg_tech3_q10_threshold_4}{18}
\pgfkeyssetvalue{mean_sentence_rating_tech3_q10}{4}
\pgfkeyssetvalue{precision_tech3_q10_threshold_5}{0.05}
\pgfkeyssetvalue{recall_tech3_q10_threshold_5}{0.25}
\pgfkeyssetvalue{total_pos_tech6_q10_threshold_5}{1}
\pgfkeyssetvalue{total_neg_tech6_q10_threshold_4}{17}
\pgfkeyssetvalue{mean_sentence_rating_tech6_q10}{4}
\pgfkeyssetvalue{precision_tech6_q10_threshold_5}{0.05}
\pgfkeyssetvalue{recall_tech6_q10_threshold_5}{0.25}
\pgfkeyssetvalue{rating_per_thread_min}{3}
\pgfkeyssetvalue{rating_per_thread_max}{10}
\pgfkeyssetvalue{rating_per_thread_median}{7}
\pgfkeyssetvalue{rating_per_thread_mode}{7}
\pgfkeyssetvalue{ground_truth_size_q8_limit_1}{76}
\pgfkeyssetvalue{neg_ground_truth_size_q8_limit_1}{0}
\pgfkeyssetvalue{total_pos_tech1_q8_threshold_1}{49}
\pgfkeyssetvalue{total_neg_tech1_q8_threshold_0}{0}
\pgfkeyssetvalue{mean_sentence_rating_tech1_q8}{3}
\pgfkeyssetvalue{precision_tech1_q8_threshold_1}{1.00}
\pgfkeyssetvalue{recall_tech1_q8_threshold_1}{0.64}
\pgfkeyssetvalue{total_pos_tech2_q8_threshold_1}{20}
\pgfkeyssetvalue{total_neg_tech2_q8_threshold_0}{0}
\pgfkeyssetvalue{mean_sentence_rating_tech2_q8}{3}
\pgfkeyssetvalue{precision_tech2_q8_threshold_1}{1.00}
\pgfkeyssetvalue{recall_tech2_q8_threshold_1}{0.26}
\pgfkeyssetvalue{total_pos_tech3_q8_threshold_1}{21}
\pgfkeyssetvalue{total_neg_tech3_q8_threshold_0}{0}
\pgfkeyssetvalue{mean_sentence_rating_tech3_q8}{3}
\pgfkeyssetvalue{precision_tech3_q8_threshold_1}{1.00}
\pgfkeyssetvalue{recall_tech3_q8_threshold_1}{0.28}
\pgfkeyssetvalue{total_pos_tech6_q8_threshold_1}{20}
\pgfkeyssetvalue{total_neg_tech6_q8_threshold_0}{0}
\pgfkeyssetvalue{mean_sentence_rating_tech6_q8}{3}
\pgfkeyssetvalue{precision_tech6_q8_threshold_1}{1.00}
\pgfkeyssetvalue{recall_tech6_q8_threshold_1}{0.26}
\pgfkeyssetvalue{ground_truth_size_q8_limit_2}{76}
\pgfkeyssetvalue{neg_ground_truth_size_q8_limit_2}{0}
\pgfkeyssetvalue{total_pos_tech1_q8_threshold_2}{49}
\pgfkeyssetvalue{total_neg_tech1_q8_threshold_1}{0}
\pgfkeyssetvalue{mean_sentence_rating_tech1_q8}{3}
\pgfkeyssetvalue{precision_tech1_q8_threshold_2}{1.00}
\pgfkeyssetvalue{recall_tech1_q8_threshold_2}{0.64}
\pgfkeyssetvalue{total_pos_tech2_q8_threshold_2}{20}
\pgfkeyssetvalue{total_neg_tech2_q8_threshold_1}{0}
\pgfkeyssetvalue{mean_sentence_rating_tech2_q8}{3}
\pgfkeyssetvalue{precision_tech2_q8_threshold_2}{1.00}
\pgfkeyssetvalue{recall_tech2_q8_threshold_2}{0.26}
\pgfkeyssetvalue{total_pos_tech3_q8_threshold_2}{21}
\pgfkeyssetvalue{total_neg_tech3_q8_threshold_1}{0}
\pgfkeyssetvalue{mean_sentence_rating_tech3_q8}{3}
\pgfkeyssetvalue{precision_tech3_q8_threshold_2}{1.00}
\pgfkeyssetvalue{recall_tech3_q8_threshold_2}{0.28}
\pgfkeyssetvalue{total_pos_tech6_q8_threshold_2}{20}
\pgfkeyssetvalue{total_neg_tech6_q8_threshold_1}{0}
\pgfkeyssetvalue{mean_sentence_rating_tech6_q8}{3}
\pgfkeyssetvalue{precision_tech6_q8_threshold_2}{1.00}
\pgfkeyssetvalue{recall_tech6_q8_threshold_2}{0.26}
\pgfkeyssetvalue{ground_truth_size_q8_limit_3}{51}
\pgfkeyssetvalue{neg_ground_truth_size_q8_limit_3}{21}
\pgfkeyssetvalue{total_pos_tech1_q8_threshold_3}{32}
\pgfkeyssetvalue{total_neg_tech1_q8_threshold_2}{14}
\pgfkeyssetvalue{mean_sentence_rating_tech1_q8}{3}
\pgfkeyssetvalue{precision_tech1_q8_threshold_3}{0.65}
\pgfkeyssetvalue{recall_tech1_q8_threshold_3}{0.63}
\pgfkeyssetvalue{total_pos_tech2_q8_threshold_3}{13}
\pgfkeyssetvalue{total_neg_tech2_q8_threshold_2}{6}
\pgfkeyssetvalue{mean_sentence_rating_tech2_q8}{3}
\pgfkeyssetvalue{precision_tech2_q8_threshold_3}{0.65}
\pgfkeyssetvalue{recall_tech2_q8_threshold_3}{0.25}
\pgfkeyssetvalue{total_pos_tech3_q8_threshold_3}{16}
\pgfkeyssetvalue{total_neg_tech3_q8_threshold_2}{4}
\pgfkeyssetvalue{mean_sentence_rating_tech3_q8}{3}
\pgfkeyssetvalue{precision_tech3_q8_threshold_3}{0.76}
\pgfkeyssetvalue{recall_tech3_q8_threshold_3}{0.31}
\pgfkeyssetvalue{total_pos_tech6_q8_threshold_3}{16}
\pgfkeyssetvalue{total_neg_tech6_q8_threshold_2}{3}
\pgfkeyssetvalue{mean_sentence_rating_tech6_q8}{3}
\pgfkeyssetvalue{precision_tech6_q8_threshold_3}{0.80}
\pgfkeyssetvalue{recall_tech6_q8_threshold_3}{0.31}
\pgfkeyssetvalue{ground_truth_size_q9_limit_1}{76}
\pgfkeyssetvalue{neg_ground_truth_size_q9_limit_1}{0}
\pgfkeyssetvalue{total_pos_tech1_q9_threshold_1}{49}
\pgfkeyssetvalue{total_neg_tech1_q9_threshold_0}{0}
\pgfkeyssetvalue{mean_sentence_rating_tech1_q9}{4}
\pgfkeyssetvalue{precision_tech1_q9_threshold_1}{1.00}
\pgfkeyssetvalue{recall_tech1_q9_threshold_1}{0.64}
\pgfkeyssetvalue{total_pos_tech2_q9_threshold_1}{20}
\pgfkeyssetvalue{total_neg_tech2_q9_threshold_0}{0}
\pgfkeyssetvalue{mean_sentence_rating_tech2_q9}{4}
\pgfkeyssetvalue{precision_tech2_q9_threshold_1}{1.00}
\pgfkeyssetvalue{recall_tech2_q9_threshold_1}{0.26}
\pgfkeyssetvalue{total_pos_tech3_q9_threshold_1}{21}
\pgfkeyssetvalue{total_neg_tech3_q9_threshold_0}{0}
\pgfkeyssetvalue{mean_sentence_rating_tech3_q9}{4}
\pgfkeyssetvalue{precision_tech3_q9_threshold_1}{1.00}
\pgfkeyssetvalue{recall_tech3_q9_threshold_1}{0.28}
\pgfkeyssetvalue{total_pos_tech6_q9_threshold_1}{20}
\pgfkeyssetvalue{total_neg_tech6_q9_threshold_0}{0}
\pgfkeyssetvalue{mean_sentence_rating_tech6_q9}{4}
\pgfkeyssetvalue{precision_tech6_q9_threshold_1}{1.00}
\pgfkeyssetvalue{recall_tech6_q9_threshold_1}{0.26}
\pgfkeyssetvalue{ground_truth_size_q9_limit_2}{74}
\pgfkeyssetvalue{neg_ground_truth_size_q9_limit_2}{1}
\pgfkeyssetvalue{total_pos_tech1_q9_threshold_2}{47}
\pgfkeyssetvalue{total_neg_tech1_q9_threshold_1}{1}
\pgfkeyssetvalue{mean_sentence_rating_tech1_q9}{4}
\pgfkeyssetvalue{precision_tech1_q9_threshold_2}{0.96}
\pgfkeyssetvalue{recall_tech1_q9_threshold_2}{0.64}
\pgfkeyssetvalue{total_pos_tech2_q9_threshold_2}{19}
\pgfkeyssetvalue{total_neg_tech2_q9_threshold_1}{1}
\pgfkeyssetvalue{mean_sentence_rating_tech2_q9}{4}
\pgfkeyssetvalue{precision_tech2_q9_threshold_2}{0.95}
\pgfkeyssetvalue{recall_tech2_q9_threshold_2}{0.26}
\pgfkeyssetvalue{total_pos_tech3_q9_threshold_2}{21}
\pgfkeyssetvalue{total_neg_tech3_q9_threshold_1}{0}
\pgfkeyssetvalue{mean_sentence_rating_tech3_q9}{4}
\pgfkeyssetvalue{precision_tech3_q9_threshold_2}{1.00}
\pgfkeyssetvalue{recall_tech3_q9_threshold_2}{0.28}
\pgfkeyssetvalue{total_pos_tech6_q9_threshold_2}{20}
\pgfkeyssetvalue{total_neg_tech6_q9_threshold_1}{0}
\pgfkeyssetvalue{mean_sentence_rating_tech6_q9}{4}
\pgfkeyssetvalue{precision_tech6_q9_threshold_2}{1.00}
\pgfkeyssetvalue{recall_tech6_q9_threshold_2}{0.27}
\pgfkeyssetvalue{ground_truth_size_q9_limit_3}{68}
\pgfkeyssetvalue{neg_ground_truth_size_q9_limit_3}{7}
\pgfkeyssetvalue{total_pos_tech1_q9_threshold_3}{44}
\pgfkeyssetvalue{total_neg_tech1_q9_threshold_2}{5}
\pgfkeyssetvalue{mean_sentence_rating_tech1_q9}{4}
\pgfkeyssetvalue{precision_tech1_q9_threshold_3}{0.90}
\pgfkeyssetvalue{recall_tech1_q9_threshold_3}{0.65}
\pgfkeyssetvalue{total_pos_tech2_q9_threshold_3}{17}
\pgfkeyssetvalue{total_neg_tech2_q9_threshold_2}{3}
\pgfkeyssetvalue{mean_sentence_rating_tech2_q9}{4}
\pgfkeyssetvalue{precision_tech2_q9_threshold_3}{0.85}
\pgfkeyssetvalue{recall_tech2_q9_threshold_3}{0.25}
\pgfkeyssetvalue{total_pos_tech3_q9_threshold_3}{19}
\pgfkeyssetvalue{total_neg_tech3_q9_threshold_2}{1}
\pgfkeyssetvalue{mean_sentence_rating_tech3_q9}{4}
\pgfkeyssetvalue{precision_tech3_q9_threshold_3}{0.90}
\pgfkeyssetvalue{recall_tech3_q9_threshold_3}{0.28}
\pgfkeyssetvalue{total_pos_tech6_q9_threshold_3}{19}
\pgfkeyssetvalue{total_neg_tech6_q9_threshold_2}{1}
\pgfkeyssetvalue{mean_sentence_rating_tech6_q9}{4}
\pgfkeyssetvalue{precision_tech6_q9_threshold_3}{0.95}
\pgfkeyssetvalue{recall_tech6_q9_threshold_3}{0.28}
\pgfkeyssetvalue{ground_truth_size_q9_limit_4}{49}
\pgfkeyssetvalue{neg_ground_truth_size_q9_limit_4}{23}
\pgfkeyssetvalue{total_pos_tech1_q9_threshold_4}{30}
\pgfkeyssetvalue{total_neg_tech1_q9_threshold_3}{16}
\pgfkeyssetvalue{mean_sentence_rating_tech1_q9}{4}
\pgfkeyssetvalue{precision_tech1_q9_threshold_4}{0.61}
\pgfkeyssetvalue{recall_tech1_q9_threshold_4}{0.61}
\pgfkeyssetvalue{total_pos_tech2_q9_threshold_4}{9}
\pgfkeyssetvalue{total_neg_tech2_q9_threshold_3}{9}
\pgfkeyssetvalue{mean_sentence_rating_tech2_q9}{4}
\pgfkeyssetvalue{precision_tech2_q9_threshold_4}{0.45}
\pgfkeyssetvalue{recall_tech2_q9_threshold_4}{0.18}
\pgfkeyssetvalue{total_pos_tech3_q9_threshold_4}{17}
\pgfkeyssetvalue{total_neg_tech3_q9_threshold_3}{4}
\pgfkeyssetvalue{mean_sentence_rating_tech3_q9}{4}
\pgfkeyssetvalue{precision_tech3_q9_threshold_4}{0.81}
\pgfkeyssetvalue{recall_tech3_q9_threshold_4}{0.35}
\pgfkeyssetvalue{total_pos_tech6_q9_threshold_4}{14}
\pgfkeyssetvalue{total_neg_tech6_q9_threshold_3}{4}
\pgfkeyssetvalue{mean_sentence_rating_tech6_q9}{4}
\pgfkeyssetvalue{precision_tech6_q9_threshold_4}{0.70}
\pgfkeyssetvalue{recall_tech6_q9_threshold_4}{0.29}
\pgfkeyssetvalue{ground_truth_size_q9_limit_5}{6}
\pgfkeyssetvalue{neg_ground_truth_size_q9_limit_5}{64}
\pgfkeyssetvalue{total_pos_tech1_q9_threshold_5}{6}
\pgfkeyssetvalue{total_neg_tech1_q9_threshold_4}{38}
\pgfkeyssetvalue{mean_sentence_rating_tech1_q9}{4}
\pgfkeyssetvalue{precision_tech1_q9_threshold_5}{0.12}
\pgfkeyssetvalue{recall_tech1_q9_threshold_5}{1.00}
\pgfkeyssetvalue{total_pos_tech2_q9_threshold_5}{3}
\pgfkeyssetvalue{total_neg_tech2_q9_threshold_4}{15}
\pgfkeyssetvalue{mean_sentence_rating_tech2_q9}{4}
\pgfkeyssetvalue{precision_tech2_q9_threshold_5}{0.15}
\pgfkeyssetvalue{recall_tech2_q9_threshold_5}{0.50}
\pgfkeyssetvalue{total_pos_tech3_q9_threshold_5}{1}
\pgfkeyssetvalue{total_neg_tech3_q9_threshold_4}{18}
\pgfkeyssetvalue{mean_sentence_rating_tech3_q9}{4}
\pgfkeyssetvalue{precision_tech3_q9_threshold_5}{0.05}
\pgfkeyssetvalue{recall_tech3_q9_threshold_5}{0.17}
\pgfkeyssetvalue{total_pos_tech6_q9_threshold_5}{1}
\pgfkeyssetvalue{total_neg_tech6_q9_threshold_4}{16}
\pgfkeyssetvalue{mean_sentence_rating_tech6_q9}{4}
\pgfkeyssetvalue{precision_tech6_q9_threshold_5}{0.05}
\pgfkeyssetvalue{recall_tech6_q9_threshold_5}{0.17}
\pgfkeyssetvalue{ground_truth_size_q10_limit_1}{76}
\pgfkeyssetvalue{neg_ground_truth_size_q10_limit_1}{0}
\pgfkeyssetvalue{total_pos_tech1_q10_threshold_1}{49}
\pgfkeyssetvalue{total_neg_tech1_q10_threshold_0}{0}
\pgfkeyssetvalue{mean_sentence_rating_tech1_q10}{4}
\pgfkeyssetvalue{precision_tech1_q10_threshold_1}{1.00}
\pgfkeyssetvalue{recall_tech1_q10_threshold_1}{0.64}
\pgfkeyssetvalue{total_pos_tech2_q10_threshold_1}{20}
\pgfkeyssetvalue{total_neg_tech2_q10_threshold_0}{0}
\pgfkeyssetvalue{mean_sentence_rating_tech2_q10}{4}
\pgfkeyssetvalue{precision_tech2_q10_threshold_1}{1.00}
\pgfkeyssetvalue{recall_tech2_q10_threshold_1}{0.26}
\pgfkeyssetvalue{total_pos_tech3_q10_threshold_1}{21}
\pgfkeyssetvalue{total_neg_tech3_q10_threshold_0}{0}
\pgfkeyssetvalue{mean_sentence_rating_tech3_q10}{4}
\pgfkeyssetvalue{precision_tech3_q10_threshold_1}{1.00}
\pgfkeyssetvalue{recall_tech3_q10_threshold_1}{0.28}
\pgfkeyssetvalue{total_pos_tech6_q10_threshold_1}{20}
\pgfkeyssetvalue{total_neg_tech6_q10_threshold_0}{0}
\pgfkeyssetvalue{mean_sentence_rating_tech6_q10}{4}
\pgfkeyssetvalue{precision_tech6_q10_threshold_1}{1.00}
\pgfkeyssetvalue{recall_tech6_q10_threshold_1}{0.26}
\pgfkeyssetvalue{ground_truth_size_q10_limit_2}{75}
\pgfkeyssetvalue{neg_ground_truth_size_q10_limit_2}{1}
\pgfkeyssetvalue{total_pos_tech1_q10_threshold_2}{48}
\pgfkeyssetvalue{total_neg_tech1_q10_threshold_1}{1}
\pgfkeyssetvalue{mean_sentence_rating_tech1_q10}{4}
\pgfkeyssetvalue{precision_tech1_q10_threshold_2}{0.98}
\pgfkeyssetvalue{recall_tech1_q10_threshold_2}{0.64}
\pgfkeyssetvalue{total_pos_tech2_q10_threshold_2}{19}
\pgfkeyssetvalue{total_neg_tech2_q10_threshold_1}{1}
\pgfkeyssetvalue{mean_sentence_rating_tech2_q10}{4}
\pgfkeyssetvalue{precision_tech2_q10_threshold_2}{0.95}
\pgfkeyssetvalue{recall_tech2_q10_threshold_2}{0.25}
\pgfkeyssetvalue{total_pos_tech3_q10_threshold_2}{21}
\pgfkeyssetvalue{total_neg_tech3_q10_threshold_1}{0}
\pgfkeyssetvalue{mean_sentence_rating_tech3_q10}{4}
\pgfkeyssetvalue{precision_tech3_q10_threshold_2}{1.00}
\pgfkeyssetvalue{recall_tech3_q10_threshold_2}{0.28}
\pgfkeyssetvalue{total_pos_tech6_q10_threshold_2}{20}
\pgfkeyssetvalue{total_neg_tech6_q10_threshold_1}{0}
\pgfkeyssetvalue{mean_sentence_rating_tech6_q10}{4}
\pgfkeyssetvalue{precision_tech6_q10_threshold_2}{1.00}
\pgfkeyssetvalue{recall_tech6_q10_threshold_2}{0.27}
\pgfkeyssetvalue{ground_truth_size_q10_limit_3}{68}
\pgfkeyssetvalue{neg_ground_truth_size_q10_limit_3}{6}
\pgfkeyssetvalue{total_pos_tech1_q10_threshold_3}{43}
\pgfkeyssetvalue{total_neg_tech1_q10_threshold_2}{5}
\pgfkeyssetvalue{mean_sentence_rating_tech1_q10}{4}
\pgfkeyssetvalue{precision_tech1_q10_threshold_3}{0.88}
\pgfkeyssetvalue{recall_tech1_q10_threshold_3}{0.63}
\pgfkeyssetvalue{total_pos_tech2_q10_threshold_3}{17}
\pgfkeyssetvalue{total_neg_tech2_q10_threshold_2}{3}
\pgfkeyssetvalue{mean_sentence_rating_tech2_q10}{4}
\pgfkeyssetvalue{precision_tech2_q10_threshold_3}{0.85}
\pgfkeyssetvalue{recall_tech2_q10_threshold_3}{0.25}
\pgfkeyssetvalue{total_pos_tech3_q10_threshold_3}{19}
\pgfkeyssetvalue{total_neg_tech3_q10_threshold_2}{1}
\pgfkeyssetvalue{mean_sentence_rating_tech3_q10}{4}
\pgfkeyssetvalue{precision_tech3_q10_threshold_3}{0.90}
\pgfkeyssetvalue{recall_tech3_q10_threshold_3}{0.28}
\pgfkeyssetvalue{total_pos_tech6_q10_threshold_3}{20}
\pgfkeyssetvalue{total_neg_tech6_q10_threshold_2}{0}
\pgfkeyssetvalue{mean_sentence_rating_tech6_q10}{4}
\pgfkeyssetvalue{precision_tech6_q10_threshold_3}{1.00}
\pgfkeyssetvalue{recall_tech6_q10_threshold_3}{0.29}
\pgfkeyssetvalue{ground_truth_size_q10_limit_4}{48}
\pgfkeyssetvalue{neg_ground_truth_size_q10_limit_4}{22}
\pgfkeyssetvalue{total_pos_tech1_q10_threshold_4}{29}
\pgfkeyssetvalue{total_neg_tech1_q10_threshold_3}{15}
\pgfkeyssetvalue{mean_sentence_rating_tech1_q10}{4}
\pgfkeyssetvalue{precision_tech1_q10_threshold_4}{0.59}
\pgfkeyssetvalue{recall_tech1_q10_threshold_4}{0.60}
\pgfkeyssetvalue{total_pos_tech2_q10_threshold_4}{11}
\pgfkeyssetvalue{total_neg_tech2_q10_threshold_3}{7}
\pgfkeyssetvalue{mean_sentence_rating_tech2_q10}{4}
\pgfkeyssetvalue{precision_tech2_q10_threshold_4}{0.55}
\pgfkeyssetvalue{recall_tech2_q10_threshold_4}{0.23}
\pgfkeyssetvalue{total_pos_tech3_q10_threshold_4}{16}
\pgfkeyssetvalue{total_neg_tech3_q10_threshold_3}{5}
\pgfkeyssetvalue{mean_sentence_rating_tech3_q10}{4}
\pgfkeyssetvalue{precision_tech3_q10_threshold_4}{0.76}
\pgfkeyssetvalue{recall_tech3_q10_threshold_4}{0.33}
\pgfkeyssetvalue{total_pos_tech6_q10_threshold_4}{16}
\pgfkeyssetvalue{total_neg_tech6_q10_threshold_3}{2}
\pgfkeyssetvalue{mean_sentence_rating_tech6_q10}{4}
\pgfkeyssetvalue{precision_tech6_q10_threshold_4}{0.80}
\pgfkeyssetvalue{recall_tech6_q10_threshold_4}{0.33}
\pgfkeyssetvalue{ground_truth_size_q10_limit_5}{4}
\pgfkeyssetvalue{neg_ground_truth_size_q10_limit_5}{64}
\pgfkeyssetvalue{total_pos_tech1_q10_threshold_5}{4}
\pgfkeyssetvalue{total_neg_tech1_q10_threshold_4}{37}
\pgfkeyssetvalue{mean_sentence_rating_tech1_q10}{4}
\pgfkeyssetvalue{precision_tech1_q10_threshold_5}{0.08}
\pgfkeyssetvalue{recall_tech1_q10_threshold_5}{1.00}
\pgfkeyssetvalue{total_pos_tech2_q10_threshold_5}{1}
\pgfkeyssetvalue{total_neg_tech2_q10_threshold_4}{16}
\pgfkeyssetvalue{mean_sentence_rating_tech2_q10}{4}
\pgfkeyssetvalue{precision_tech2_q10_threshold_5}{0.05}
\pgfkeyssetvalue{recall_tech2_q10_threshold_5}{0.25}
\pgfkeyssetvalue{total_pos_tech3_q10_threshold_5}{1}
\pgfkeyssetvalue{total_neg_tech3_q10_threshold_4}{18}
\pgfkeyssetvalue{mean_sentence_rating_tech3_q10}{4}
\pgfkeyssetvalue{precision_tech3_q10_threshold_5}{0.05}
\pgfkeyssetvalue{recall_tech3_q10_threshold_5}{0.25}
\pgfkeyssetvalue{total_pos_tech6_q10_threshold_5}{1}
\pgfkeyssetvalue{total_neg_tech6_q10_threshold_4}{17}
\pgfkeyssetvalue{mean_sentence_rating_tech6_q10}{4}
\pgfkeyssetvalue{precision_tech6_q10_threshold_5}{0.05}
\pgfkeyssetvalue{recall_tech6_q10_threshold_5}{0.25}

\begin{document}
\title{Essential Sentences for Navigating Stack Overflow Answers}

\author{
\IEEEauthorblockN{Sarah Nadi}
\IEEEauthorblockA{
\textit{University of Alberta}\\
Edmonton, AB, Canada \\
nadi@ualberta.ca}
\and
\IEEEauthorblockN{Christoph Treude}
\IEEEauthorblockA{
\textit{University of Adelaide}\\
Adelaide, SA, Australia \\
christoph.treude@adelaide.edu.au}
}

\maketitle

\begin{abstract}
Stack Overflow (SO) has become an essential resource for software development.
Despite its success and prevalence, navigating SO remains a challenge. 
Ideally, SO users could benefit from highlighted \navcue{s} that help them decide if an answer is relevant to their task and context.
Such \navcue{s} could be in the form of \textit{essential sentences} that help the searcher decide whether they want to read the answer or skip over it.
In this paper, we compare four potential approaches for identifying essential sentences.
We adopt two existing approaches and develop two new approaches based on the idea that contextual information in a sentence (e.g., ``if using windows'') could help identify essential sentences.
We compare the four techniques using a survey of \pgfkeysvalueof{final_num_participants} participants.
Our participants indicate that it is not always easy to figure out what the best solution for their specific problem is, given the options, and that they would indeed like to easily spot contextual information that may narrow down the search.
Our quantitative comparison of the techniques shows that there is no single technique sufficient for identifying essential sentences that can serve as \navcue{s}, while our qualitative analysis shows that participants valued explanations and specific conditions, and did not value filler sentences or speculations.
Our work sheds light on the importance of navigational cues, and our findings can be used to guide future research to find the best combination of techniques to identify such cues.

\end{abstract}

\section{Introduction}

With more than 18 million questions and 28 million answers as of October 2019, Stack Overflow (SO) has become a critical knowledge asset for the software development industry. However, navigating the amount of data available on Stack Overflow is challenging.
While lots of previous work focused on identifying the most relevant threads for a given query~\cite{Ponzanelli2014, deSouza2014, RahmanICSME18,Silva:2019}, navigation is not complete after a suitable thread has been identified.
In a survey with 72 developers from two IT companies, Xu et al.~\cite{Xu2017} identified ``the answers in long posts are hard to find'' as one of the challenges. Almost 6.5 million questions (37\% of all questions) have more than one answer, and the average length of an answer is 789 characters. Identifying useful information from this amount of data is not trivial.
Thus, in this paper, we focus on the next stage of the navigation process.
Specifically, we assume that the user has already identified the relevant threads and now needs to navigate them to identify relevant information.

\begin{figure}
\centering
\includegraphics[width=0.5\textwidth]{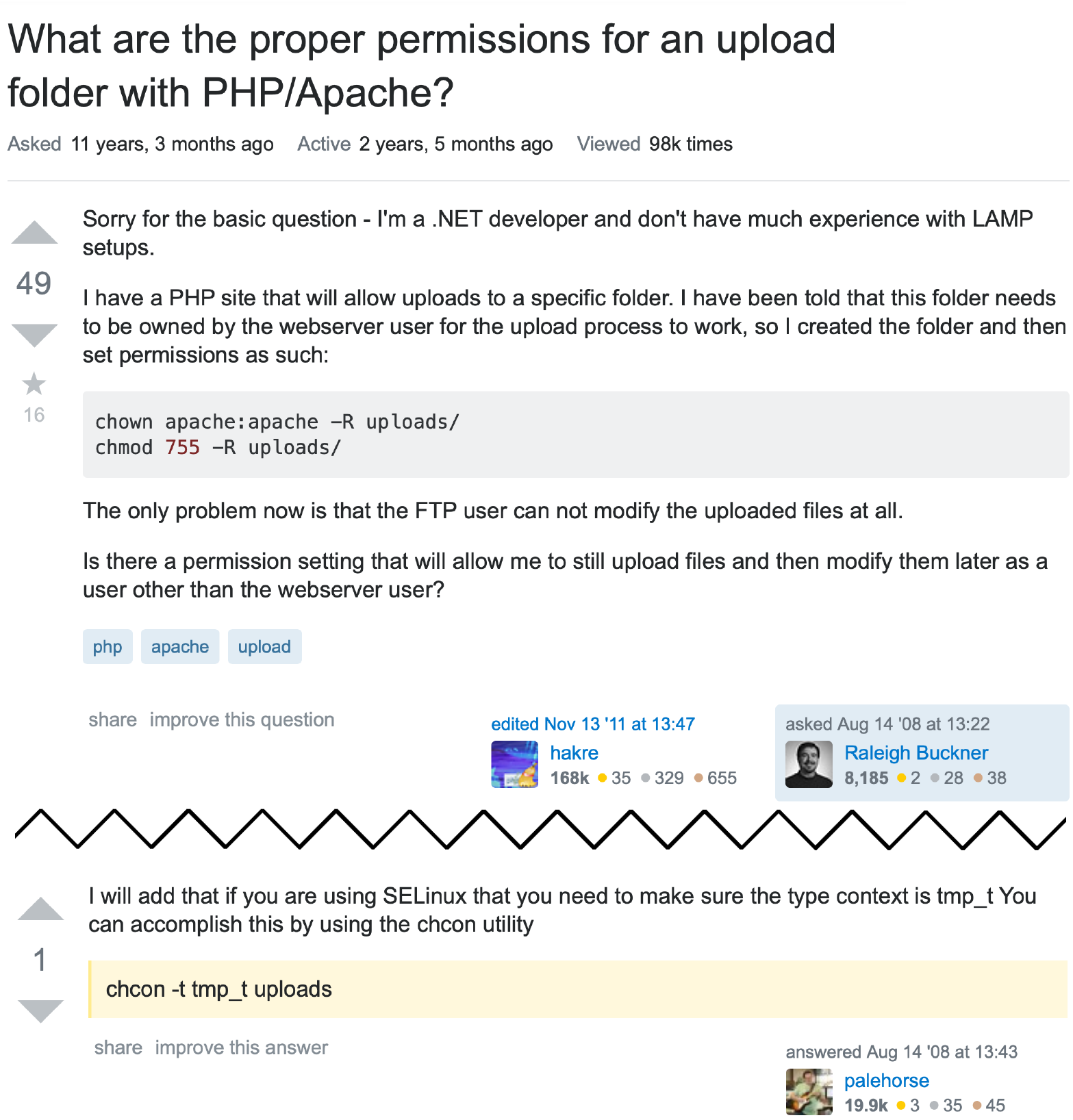}
\caption{Motivating Example from SO Thread 10990. The sentence mentioning SELinux is an essential sentence for navigating this thread. Our work compares four approaches for automatically identifying such sentences.}
\label{fig:motivating}
\vspace{-0.6cm}
\end{figure}

Stack Overflow threads that contain many and/or long answers often discuss alternatives or opinions~\cite{Uddin2019, Zagalsky2018,Lin:2019}.
A user browsing such threads may want to quickly spot certain information about an answer to decide if it fits their needs.
We refer to this information as \textit{essential sentences}, because clearly seeing these sentences can help the user quickly decide if they should skip over this answer or carefully read it.
The question is how can we identify such essential sentences?

We could think of two possibilities from existing work.
The first, \wordpattern, is a pattern-based approach used to identify indispensable information in software documentation~\cite{Robillard2015}.
The second is \lexrank, which is a standard text summarization algorithm that identifies the most important sentence(s) in a document. 
If we assume that the developer has a certain context in mind when navigating information (e.g., a particular platform, technology stack, or non-functional requirement), then another possibility for identifying essential sentences is looking for conditional sentences. 
For example, Figure~\ref{fig:motivating} shows a screenshot of Stack Overflow question \post{10990} along with one of its answers~\answer{11025}. At the time of writing, this answer appears as the sixth out of eight answers in Stack Overflow's default ordering of answers by votes, i.e., the user would have to scroll past five other answers, including an accepted one, to see the sentence which indicates that this answer might be relevant to their particular context: For users using Security-Enhanced Linux (SELinux), it is important to know that type context needs to be \texttt{tmp\_t} as part of setting the permissions for an upload folder.
Thus, \textit{``I will add that if you are using SELinux...''} is an essential sentence which if highlighted could help with navigating the information.

In this paper, we compare four approaches for identifying essential sentences in Stack Overflow answers.
The first two are \wordpattern and \lexrank mentioned above.
The other two approaches are both built on the above intuition that contextual information is often expressed as a condition in the sentence.
Thus, the third approach, \simpleif, simply relies on identifying  any sentence with an if condition, or in other words \textit{conditional sentences}.
The fourth approach \condinsight aims to identify the subset of conditional sentences that specifically express a relevant condition or that provide contextual information, such as ``\textit{..if you are using SELinux}'' from Figure~\ref{fig:motivating}.
We evaluate \pgfkeysvalueof{total_sentence_eval} sentences identified by the four approaches through a survey with \pgfkeysvalueof{final_num_participants} participants. Our survey answers the following research questions:

\begin{enumerate}[label=\subscript{RQ}{\arabic*},leftmargin=*]
\item  \label{rq1} What navigation challenges do SO users face? 
\item \label{rq2} Can highlighting navigational cues help SO users? 
\item \label{rq3} Are essential sentences identified by different approaches perceived as helpful \navcue{s}?
\item \label{rq4} What information is available in essential sentences?
\end{enumerate}

Our participants indicate that sifting through information on Stack Overflow to find the most suitable answer they can adapt to their specific context is a challenge. The results also suggest that highlighting relevant information, including contextual information, may help.
Our quantitative results show that while \lexrank has the highest percentage of highly rated sentences, each technique finds a different set of highly rated sentences.
This suggests that there is no single technique that is clearly suitable for identifying essential sentences.
However, our qualitative analysis shows that the majority of positively rated essential sentences contained explanations or specific conditions, which can help future work create more techniques for identifying essential sentences for navigation cues.

\section{Related Work}

To the best of our knowledge, there is no existing work on the extraction of essential sentences to be used as \navcue{s}. However, there is a lot of relevant work on identifying various types of information in Stack Overflow or software documentation, which we review in the following.

\paragraph{Identifying relevant information on Stack Overflow}

Gottipati et al.~\cite{Gottipati2011} argue that it is hard for developers to find relevant answers in question-and-answer forums. They introduced a semantic search engine to recover relevant answers. The motivation of our work follows a similar thought, but at a finer level of granularity: Not only is it hard to find relevant answers on SO, but it is also challenging to navigate the information in these answers. Nasehi et al.~\cite{Nasehi2012} found that explanations accompanying SO code examples are as important as the examples themselves, further motivating our work on helping users navigate these textual explanations to identify relevant information.

Ye et al.~\cite{Ye2014} and Zou et al.~\cite{Zou2015} investigated the types of interrogatives (e.g., how-to, what, and where) in questions and answers and leverage this information to re-rank search results.
Bagheri and Ensan~\cite{Bagheri2016} propose a semantic tagging approach based on Wikipedia data to improve Stack Overflow tags, eventually improving the search process. 
Soliman et al.'s~\cite{Soliman2018} approach on the other hand is more domain-specific, focusing on how architects search for architecturally relevant information on Stack Overflow. 
CROKAGE by Silva et al.~\cite{Silva2019} takes the description of a programming task and provides a comprehensive solution for this task by searching multiple threads.
In contrast, our work focuses on the navigation of a single thread, with the goal of identifying \navcue{s}. 
Finally, Opiner~\cite{Uddin2017} and POME~\cite{Lin:2019} identify Stack Overflow sentences that specifically contain opinions about Application Programming Interfaces (APIs), with the end-goal of recommending an API that matches a specific aspect (e.g., performance).
In our work, we focus on sentences that help developers navigate answers regardless of whether they contain a positive or negative sentiment. 

\paragraph{Identifying relevant information in software documentation}

In an approach to bridge documentation from different sources, Treude and Robillard~\cite{Treude2016} augmented API documentation with insight sentences from Stack Overflow, i.e., sentences that are related to a particular API type and that provide insights not contained in the API documentation of that type. While some insight sentences may also be navigational cues, the application is fundamentally different: we focus on highlighting \navcue{s} in Stack Overflow threads instead of extracting them to complement a different information source; we do not rely on API documentation to identify sentences.
However, given the somewhat similar goals, we use their same baselines.
Specifically, we use the \wordpattern approach by Robillard and Chhetri~\cite{Robillard2015}, who argued that information useful to programmers can be buried in irrelevant text, making it difficult to discover. To overcome that, they detected fragments of API documentation potentially important to a programmer, based on a tool which supports both information filtering and discovery.
We also use the same \lexrank summarization technique they compare to. 

Outside of SO, Petrosyan et al.~\cite{Petrosyan2015} use supervised text classification based on linguistic and structural features to discover tutorial sections explaining a given API type. Their approach uses supervised text classification. Jiang et al.~\cite{Jiang2016} presented a similar approach, discovering two important indicators to complement traditional text-based features, namely co-occurring APIs and knowledge-based API extensions. They later refined their work~\cite{Jiang2017} by identifying APIs in tutorial fragments and replacing ambiguous pronouns and variables with related ontologies and API names.
Specifically focusing on the challenge of navigating software documentation, Treude et al.~\cite{Treude2015} automatically extracted tasks from software documentation and suggested them to developers.
Tian et al.~\cite{Tian2017} developed a bot to answer API questions given API documentation as an input. Zhong et al.~\cite{Zhong2009} inferred specifications from API documentation by detecting actions and resources through machine learning. 

Different to these approaches which are generally aimed at more structured documentation, such as API specifications, we focus on identifying \navcue{s} in SO threads.

\section{Identifying Essential Sentences}
\label{sec:techniques}

In this paper, we compare four techniques for identifying essential sentences in Stack Overflow answers. Essential sentences can be used as \navcue{s} that allow a user to easily find the information they are looking for.

The first two techniques we use are \wordpattern and \lexrank, which have been used as the baselines in previous work on identifying insight sentences for API documentation~\cite{Treude2016}.
Since both techniques try to identify various forms of valuable information, which may in turn be relevant for navigating Stack Overflow threads, we evaluate their application in our context of identifying essential sentences.

We also develop two new techniques that rely on the concept of conditions in sentences.
The first, \simpleif, simply identifies all sentences with the word ``if'', regardless of the contents of the condition.
The second technique, \condinsight, finds conditional sentences whose conditional phrase contains relevant technical context.
We develop several heuristics to determine whether the condition in the sentence is useful.
We now present the details of all four techniques, first discussing 
the common pre-processing steps we apply for all of them.

\subsection{Common Pre-processing Steps}
\label{sec:preproces}

Given a SO thread, we use \texttt{BeautifulSoup}~\cite{bsoup} to process each answer's html code.
We identify all paragraphs by searching for \texttt{<p>} tags.
For each paragraph, we replace all html links (identified by the \texttt{<a>} tag) with the word \texttt{LINK} to enable the processing of the sentence through natural language processing tools later.
For similar reasons, we also replace all in-text code (identified by the \texttt{$<$code$>$} tag) with the word \texttt{CW}.
Afterwards, we use the Stanford CoreNLP toolkit~\cite{corenlp} to identify the sentences in each paragraph.
All techniques work on the same set of identified sentences. 
When parsing a paragraph using CoreNLP, we use the part-of-speech tagger (pos) and parse annotator, which provide full syntactic analysis using both the constituent and the dependency representations. Such annotators extract the tree structure of the sentence, allowing us to later identify any conditional phrases.

\subsection{Technique 1: \wordpattern} 

As a potential technique for identifying essential sentences, we use the work by Robillard and Chhetri~\cite{Robillard2015}, who developed a set of 360 word patterns for identifying sentences containing \textit{indispensable} knowledge in documentation pages.
The word patterns they created typically consist of a set of words and a code word that must appear in a sentence. An example of such a word pattern is \{should, value, CW, be\}.
This pattern indicates that these three words must appear in the sentence, as well as any code word.
An example matching this pattern is \textit{``Providing the \texttt{contentType} parameter with the value of \texttt{json} will tell jQuery that the response should be json and it should auto parse it before giving it to a callback.''}

Given a sentence, as identified through the common pre-processing steps,
we lemmatize all words in the sentence to increase the chances of a pattern match.
Since some users may use code words such as a class or a function name in their text without necessarily formatting it as code using the \texttt{$<$code$>$} html tag, we also use the list of regular expressions from Treude and Robillard's work~\cite{Treude2016} to identify additional code elements in the sentence and replace them with \texttt{CW} to further increase the chance of matching code words.
We then look for the existence of any word pattern by searching for \textit{all} the words in the pattern list, including \texttt{CW}. 
If the sentence matches any of the given 360 patterns, we mark this sentence as a \texttt{wordpatttern} sentence.

\subsection{Technique 2: \lexrank}

Text summarization identifies the most important sentences from a given document~\cite{Erkan2004}.
We could think of \navcue{s} as sentences that contain important information relevant to their surrounding sentences, and thus interpret the task as a text summarization task.
Lexrank~\cite{Erkan2004} is a commonly used unsupervised text summarization approach.
We use \lexrank here as another potential technique for identifying \navcue{s} in a thread.
We apply the common pre-processing steps and then lemmatize all words in the sentences.
Finally, we pass the whole pre-processed thread to an existing open-source implementation of lexrank~\cite{lexrankimpl}, and indicate the number of sentences it should return (1 sentence, see Section~\ref{sec:thread-sel}). 
For example, for thread \post{50957609}, lexrank identifies the sentence ``\textit{You're not looping over an Array, you are looping over the properties of an Object with a for...in loop}''.

\subsection{Technique 3: \simpleif} 

Essential sentences may contain contextual information which may often be expressed in the form of conditions, such as that shown in the example in the introduction.
Thus, as a third potential technique for identifying essential sentences, we propose \simpleif, which identifies all sentences that simply have the word ``if'' in them and thus contain a conditional phrase. 
We apply the common pre-processing steps and then keep sentences with the word ``if'' in them.
An example \simpleif sentence from thread \post{52853048} is \textit{``If not, you could insert an extra trailing value, e.g. null, if it doesn't hurt.''}

\subsection{Technique 4: \condinsight}
\label{sec:condinsight}

\simpleif finds all conditional sentences.
In \condinsight, the goal is to automatically identify conditional sentences that carry technical context and are useful.
We do this based on a set of heuristics that we identified through manually analyzing 118 randomly selected conditional sentences from our corpus.

\paragraph*{(1) Conditional Phrase Extraction} Given a \simpleif sentence, we extract the conditional phrase from the parse tree produced by CoreNLP.
Specifically, we look for subordinate clauses (SBAR) with an ``if'' in the left subtree, and then extract the conditional phrase from the simple declarative clause that is in the corresponding right subtree.
For example, the conditional phrase of \textit{``...., which will outperform the sync method if your server is under load''} is ``\textit{if your server is under load}.’’
We then identify all nouns in the conditional phrase, \textit{load} and \textit{server} in the example sentence.
We also treat any code elements in the sentence, again identified by the regular expressions from Treude and Robillard~\cite{Treude2016}, as nouns.
We then compare this noun list to the set of all Stack Overflow tags, with the goal of ensuring that the condition has some technical context.
If any of the nouns are also SO tags, then we proceed to the next steps.
Otherwise, we discard this sentence.
In our example sentence, both nouns are also SO tags.

\paragraph*{(2) Grammatical-relationship filtering} Based on our manual analysis, we noticed that the sentence's grammar structure can often be an indication of its usefulness. We thus follow intuition from Wang et al.~\cite{wong2013autocomment} about using grammatical dependencies to identify higher quality sentences from Stack Overflow. We keep only sentences whose conditional phrase (1) has a verb phrase where the verb has a dependency on a noun (e.g., \textit{if you want a good UI}) or (2) where the ``if'' has a direct dependency on a noun (e.g., \textit{if file exists}). 

\paragraph*{(3) Final heuristic filtering} 

As a final step, we further remove sentences: ending in a question mark, with first person or ``you'' references but no modal verb (e.g., ``if you look at the data''), containing unsure phrases (e.g., \textit{I'm not sure if that's...}) or with conditions surrounded by parentheses. 
An example \condinsight sentence from thread \post{52703976} is \textit{``If your expected value is an array, consider using map, especially if there will always be a value''}.

\section{Evaluation Setup}
In this section, we describe the setup we use to evaluate the sentences identified by each of the four techniques.

\subsection{Thread Selection}
\label{sec:thread-sel}

We select \texttt{json} as our subject domain, because it is a general data exchange format that is used in multiple programming languages and technologies.
This increases our chance to get meaningful ratings, as opposed to using a specific technology that only few participants would be familiar with.
We use the StackExchange API to find \texttt{json}-tagged threads that (1) have a question score of zero or more to filter out questions that have been explicitly marked as negative
and (2) have been asked between March 29, 2018 to March 29, 2019.

The search returned 29,420 threads.
We consider only threads that have a minimum of two answers, to ensure there are at least two alternative answers such that highlighting essential sentences makes sense.
This leaves us with 7,920 threads.
We processed 19,427 answers for these threads, with a mean of 2.45 answers per thread, and some threads having up to 13 answers.
We run all techniques on all text from these answers, processing a total of 68,331 sentences.
In the end, we identified 1,200 \wordpattern sentences, 3,441 \simpleif sentences, and 761 \condinsight sentences.

Since evaluating thousands of sentences in a survey is infeasible, we sample the threads for evaluation.
To avoid being biased towards one technique, our criteria for selecting a thread for evaluation is that all techniques detected at least one sentence in that thread.
\texttt{lexrank} needs the number of sentences to select for a summary, which guarantees that it will always identify at least one sentence per thread.
Thus, we look only at the results of \wordpattern, \condinsight, and \simpleif for our sample selection.
We identified 79 threads for which all three techniques detected at least one sentence.
From these, we randomly select 20 threads for evaluation.
To avoid bias, we want to balance the number of sentences being evaluated for each technique.
Therefore, we first randomly select 20 threads.
We then look at the number of total sentences identified from each of the three techniques in those 20 threads.
If there is a big imbalance (e.g., a thread with 6 \wordpattern sentences vs. 3 \condinsight and 3 \simpleif or a thread with 5 \simpleif sentences and only one sentence from each of the other techniques), we replace the thread with another randomly selected one until we get a reasonable and balanced number of sentences in each technique.
To avoid over-burdening participants, we also focus on threads with at most a total of 5 sentences selected by all approaches.
To determine the number of sentences to provide to the \texttt{lexrank} algorithm, we use the median number of sentences detected by each technique for these 20 threads.
Since this median is 1 sentence, we configure \texttt{lexrank} to select one sentence per thread.

\begin{figure}[t!]%
\centering
\includegraphics[width=0.3\textwidth]{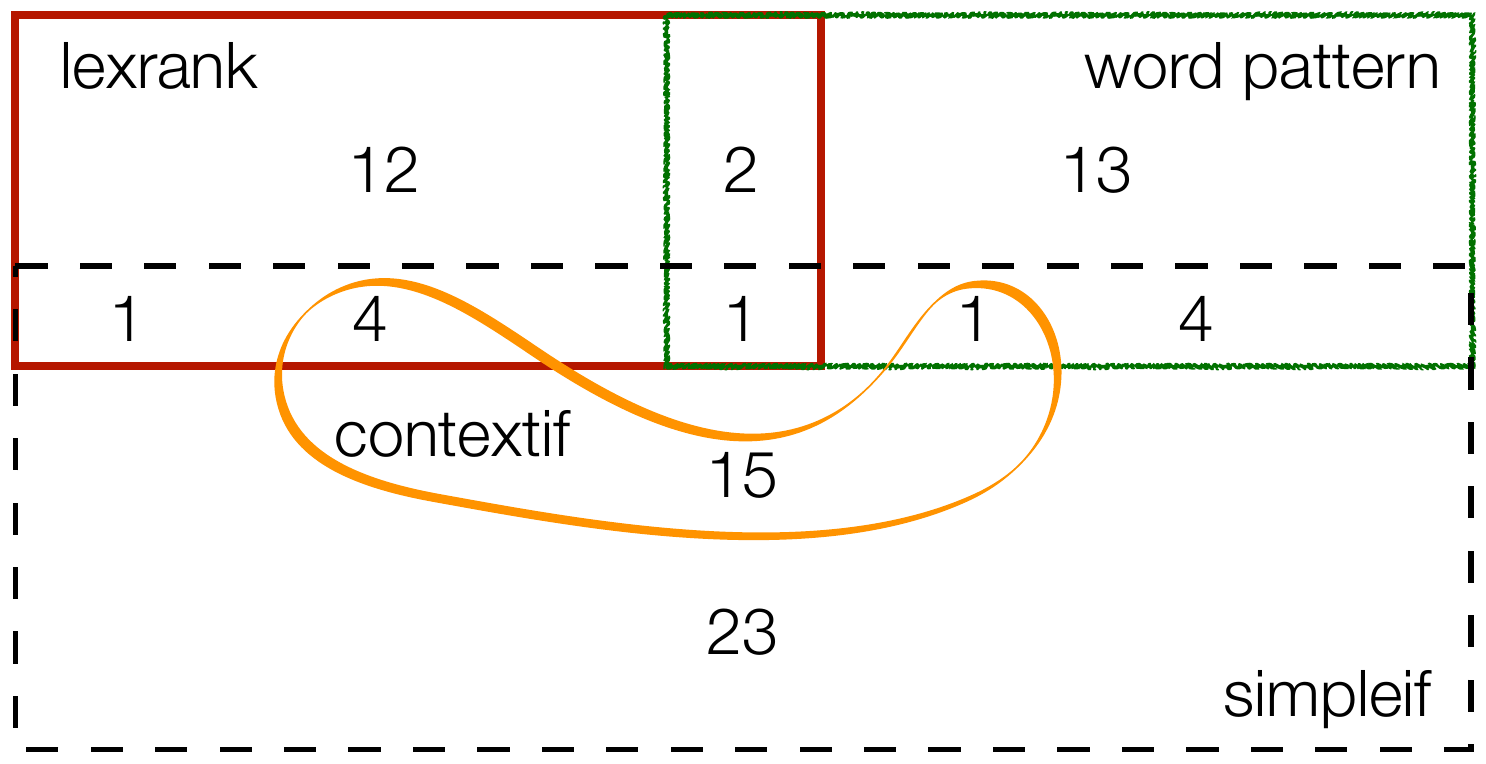}
\caption{Venn diagram illustrating the overlap between the \pgfkeysvalueof{total_sentence_eval} sentences identified across the four techniques}%
\vspace{-0.5cm}
\label{fig:sentences_overlap}	
\end{figure}

Our final 20 selected threads, along with their descriptive statistics, are available on our artifact page~\cite{artifact}.
The minimum number of answers per thread was 2, median 3, and maximum 7.
Table~\ref{tab:selectedsentences} shows the number of sentences identified by each technique across the 20 selected threads.
Note that the same sentence may be identified by multiple techniques.
In total, we have \pgfkeysvalueof{total_sentence_eval} unique sentences for evaluation in the survey.
As shown in Figure~\ref{fig:sentences_overlap}, only 13 out of the  \pgfkeysvalueof{total_sentence_eval} unique sentences are identified by more than one technique.
Furthermore, the majority of sentences identified by each category of techniques are unique to that technique, suggesting that these techniques detect different kinds of information.

\subsection{Survey Design}

We design a custom web application that shows highlighted sentences within the context and GUI of a Stack Overflow thread.
Participants see a page that contains the full thread, but without the extra controls/buttons/information on SO pages to reduce unrelated clutter.
When they start the survey, participants are provided with an information page that describes the study (including ethics clearance information) and then proceed to an instructions page that explains their task.
We now describe the flow of the survey.

\paragraph{Background Questions}

Participants first see the background questions page, with the following questions.

\begin{enumerate}[label=(\subscript{BQ}{\arabic*}),leftmargin=*]
\item \textit{Is developing software part of your job?} Yes, no

\item \textit{What is your job title?} Free text

\item \textit{For how many years have you been developing software?} Free text

\item \textit{What is your area of software development?} Free text

\item \label{q:jsonbg} \textit{How would you rate your json expertise?} No experience at all using json, Beginner, Intermediate, Expert

\item \label{q:sosearch} \textit{Have you used Stack Overflow to search for information before?} Yes, no

\item \textit{Have you contributed to Stack Overflow before? (questions, answers, comments, discussion etc.)} Yes, no
\end{enumerate} 

\begin{figure}[t!]
\centering
\includegraphics[width=0.47\textwidth]{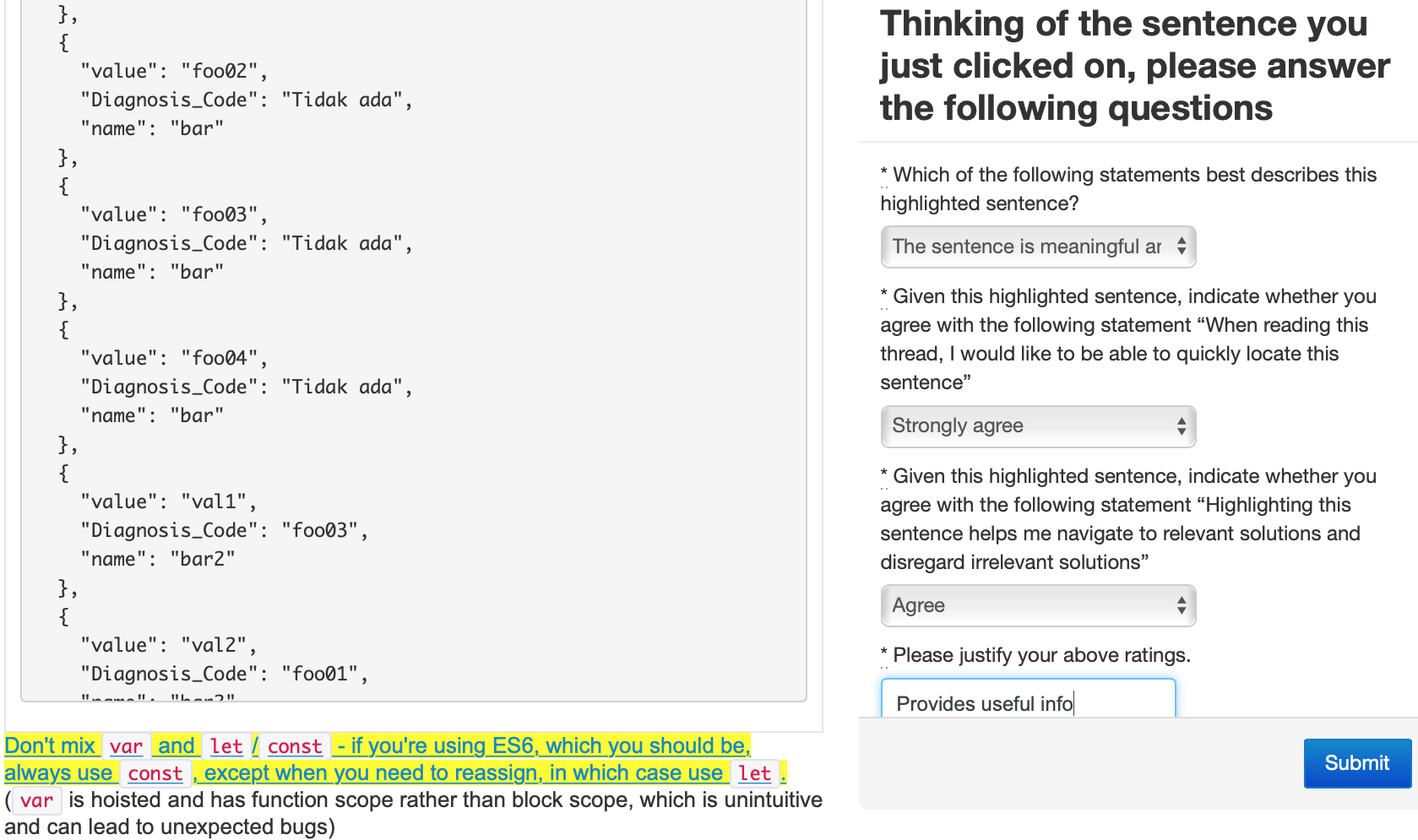}
\caption{Sentence questions displayed to survey participants}%
\vspace{-0.5cm}
\label{fig:survey-snapshot}
\end{figure}

\paragraph{Sentence Review Questions}
After answering the background questions, participants proceed to review three threads plus one quality gate thread (explained shortly).
The sentences identified by the four techniques are highlighted in each thread.
All sentences are highlighted in the same color and format, regardless of the technique that detected them to avoid any bias.
Thus, each participant evaluates sentences from all four techniques without actually knowing which technique they are evaluating.
Once participants click on a highlighted sentence, the following questions appear in the right margin beside the sentence (as shown in Figure~\ref{fig:survey-snapshot}).

\begin{enumerate}[noitemsep,label=(\subscript{SR}{\arabic*}),leftmargin=*]
\item \label{q:q8} \textit{Which of the following statements best describes this highlighted sentence?}
(a) The sentence is meaningful and provides important/useful information needed to correctly accomplish the task in question\footnote{Refers to the question being asked in the current SO thread}, 
(b) The sentence is meaningful, but does not provide any important/useful information to correctly accomplish the task in question,
(c) The sentence does not make sense to me. Note that we ask this question to differentiate wording/grammar issues from actual usefulness. 

\item \label{q:q9} \textit{Given this highlighted sentence, indicate whether you agree with the following statement ``When reading this thread, I would like to be able to quickly locate this sentence''} (a) strongly agree, (b) agree, (c) neither agree or disagree, (d) disagree, (e) strongly disagree.

\item \label{q:q10} \textit{Given this highlighted sentence, indicate whether you agree with the following statement ``Highlighting this sentence helps me navigate to relevant solutions and disregard irrelevant solutions''} (a) strongly agree, (b) agree, (c) neither agree or disagree, (d) disagree, (e) strongly disagree.
\item \label{q:q11} \textit{Please justify your above ratings.} Free text.
\end{enumerate} 

As a \textit{quality gate}, we fix one thread that we selected beforehand with the following sentence highlighted: \textit{``Hope this helps.''} 
We expect participants who are not randomly answering the questions to rate this sentence negatively since it does not provide any useful information.
The quality gate thread appears in a random order for each participant and looks the same as any of the other threads.

\paragraph{Exit Questions}

After evaluating four threads, participants proceed to the following final set of exit questions.

\begin{enumerate}[noitemsep,label=(\subscript{EQ}{\arabic*}),leftmargin=*]
\item \label{q:q12}\textit{Thinking of the highlighted sentences you observed, and specifically the ones you thought were useful/important, are there any specific properties of these sentences that affected your rating?}  Free form
\item \label{q:q13}\textit{Thinking of the highlighted sentences you observed, and specifically the ones you thought were NOT useful/important, are there any specific properties of these sentences that affected your rating?} Free form
\item \label{q:q14}\textit{Thinking of your general experience as a software developer: after you have identified a relevant Stack Overflow thread that you need to look at, what challenges (if any) do you encounter in navigating the information in the thread?} Free form
\item \label{q:q15}\textit{Assuming Stack Overflow could highlight certain information for you in a given thread, what kind of information would you like to see?} Free form
\item \label{q:q16} \textit{Some of the sentences you saw were conditional sentences (i.e., sentences with an if clause) that contained conditions related to programming languages, technologies, operating systems, or situations a developer would face. Do you think highlighting such sentences could be useful to find relevant information in a thread more quickly?} Free form

\end{enumerate} 

While the threads each participant sees are randomly selected from the thread pool and appear in a random order, we use a balancing algorithm to ensure that we get at least 3 ratings per thread.
Thus, our algorithm tries to select from the threads with the lowest number of responses first.

\subsection{Participant Recruitment}

We used Amazon Mechanical Turk~\cite{mechturk} (MT) to recruit participants, using the premium option of ``Employment Industry -- Software \& IT Services'' as a required qualification and a compensation of USD\$3.
Our web application generates a unique token for each participant to submit on MT. 
We accepted only responses which provide this token and where the participant has (at least) minimal knowledge of JSON~\ref{q:jsonbg} and has used SO to search for information before~\ref{q:sosearch}.
Each participant can answer the survey only once.

We also performed post-filtering of the responses to ensure that we consider results only from participants who understood the task.
We use our quality gate thread to filter out participants.
Specifically, we look at the answers to \ref{q:q10} for the sentence ``\textit{Hope this helps.}''
 and filter out all participants who did not answer ``Disagree'' or ``Strongly disagree''.

To ensure that there are no issues with our survey application, we first conducted a pilot study with 5 MT participants.
We fixed some technical issues that arose from the pilot.
We do not include the pilot data in the results presented in this paper, and only base our results on the full run.

\begin{figure}[t!]%
\centering

	\begin{subfigure}[t]{0.4\textwidth}%
		\centering
		\includegraphics[width=\textwidth]{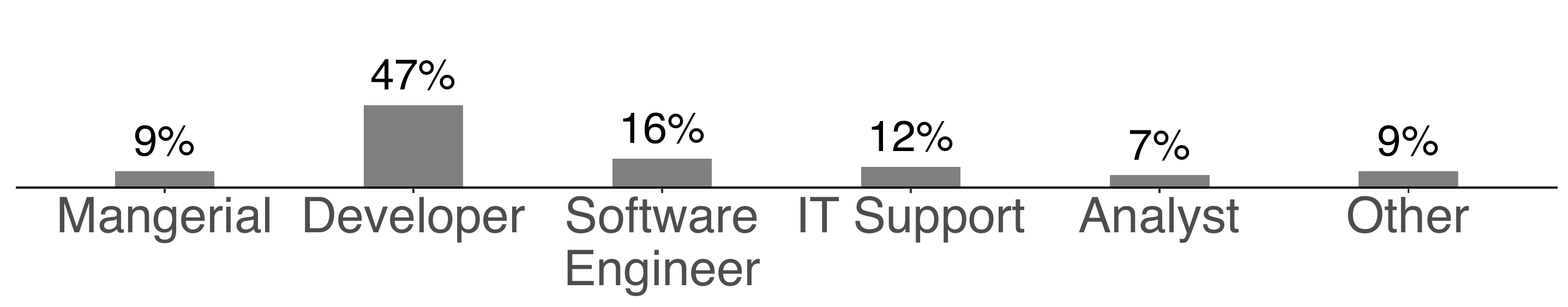}
		\caption{Job title}%
		\label{fig:jobtitle}	
	\end{subfigure}
	\begin{subfigure}[t]{0.4\textwidth}%
		\centering
		\includegraphics[width=\textwidth]{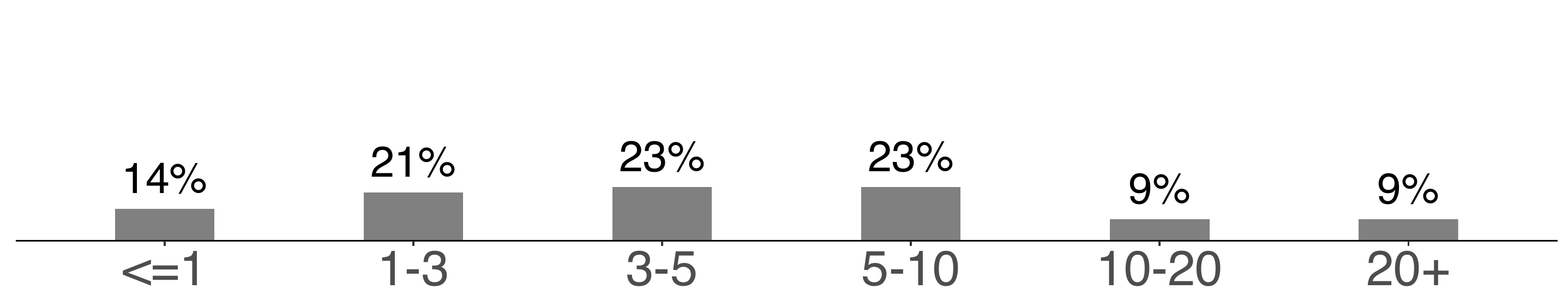}
		\caption{Years of software development}%
		\label{fig:dev-years}	
	\end{subfigure}

	\begin{subfigure}[t]{0.4\textwidth}%
		\centering
		\includegraphics[width=\textwidth]{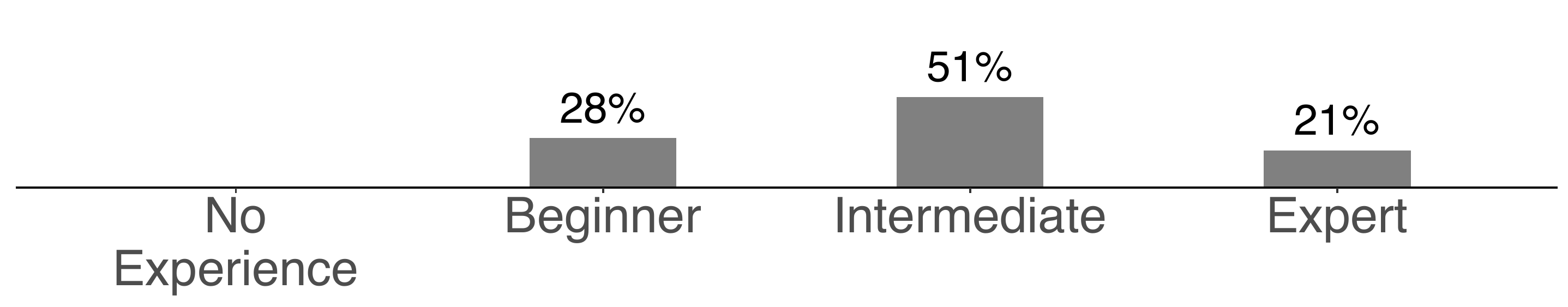}
		\caption{What is your experience using json?}%
		\label{fig:json-exp}	
	\end{subfigure}

	\caption{Participant Background}
		\label{fig:partic_bg}
		\vspace{-0.3cm}
\end{figure}

\subsection{Data Analysis}

\paragraph*{Qualitative Analysis}
For RQs 1, 2, and 4, we use qualitative methods, specifically open coding~\cite{Stol2016}, to analyze participants' responses to the exit questions, as well as the information contained in positively rated sentences.
For each analysis, one author first manually created meaningful codes~\cite{Stol2016} for a random sample of the data.
Then, using these defined codes, both authors coded~10\% of the data points using the defined coding scheme.
If our inter-rater reliability, calculated using Fuzzy Kappa~\cite{Kirilenko2016} to account for potentially multiple labels per data point, was high ($>= 0.75$), the two authors continued coding the rest of the data.
If not, they discussed the codes first and found potentials for code merging.

\paragraph*{Quantitative Analysis} To answer RQ2, we need to compare the ratings of~\ref{q:q8},~\ref{q:q9}, and~\ref{q:q10} for the different techniques.
A high median rating suggests that the majority of participants thought that this sentence is ``good'' according to the criteria the question asks about.
Thus, to evaluate the effectiveness of each technique, we look at the percentage of sentences identified by that technique which received a high rating.
We use the question nature to determine what a ``high rating'' means. 
For Question~\ref{q:q8}, we convert the choices to ratings, where 1 corresponds to not meaningful, 2 corresponds to meaningful but not useful/helpful, and 3 corresponds to meaningful and useful/helpful.
Since we are interested in meaningful and helpful sentences, we consider a sentence with a median score equal to 3 as a highly rated sentence.
For Questions~\ref{q:q9} and~\ref{q:q10}, we convert the choices to ratings from 1 (Strongly disagree) to 5 (Strongly agree). 
For those questions, we are interested in sentences where participants at least agree with the statement.
Thus, we consider sentences with median score $>=4$ as highly rated. 

\section{Survey Summary Statistics}

We accepted submissions from \pgfkeysvalueof{original_participants} participants.
We filter out \pgfkeysvalueof{filtered_out_participants} participants based on our quality gate thread, and use the remaining \pgfkeysvalueof{final_num_participants} participants for our results.

\Figref{fig:partic_bg} shows the distribution of the background of the \pgfkeysvalueof{final_num_participants} survey participants.
We group related job titles under common categories.
As Figures~\ref{fig:jobtitle} and~\ref{fig:dev-years} show, our participants work in various occupations related to the technology sector and have a wide range of years of software development experience.
\Figref{fig:json-exp} shows that the majority of our participants had intermediate knowledge of json; recall that we did not accept submissions from participants with no json experience. 

As per our load balancing algorithm, all threads received at least \pgfkeysvalueof{rating_per_thread_min} ratings.
The median number of ratings per thread was \pgfkeysvalueof{rating_per_thread_median}, with a maximum of \pgfkeysvalueof{rating_per_thread_max}.
The median number of ratings per sentence per technique was also \pgfkeysvalueof{rating_per_thread_median}, and the distribution of the number of ratings per sentence across the four techniques was similar, giving us confidence that the number of ratings will not implicitly bias our results towards a certain technique.

\section{RQ1: Navigation Challenges} 

In \ref{rq1}, we want to understand what kind of navigation challenges users face, if any. We look at the SO navigation challenges participants mention in~\ref{q:q14}. 
The codes we use for these responses, along with the number of responses per code, are: \textit{too much information to navigate} (13), \textit{no problems} (11), \textit{not easy to adapt information to my specific problem} (6), \textit{outdated information} (5), \textit{SO feature} (2), \textit{need to combine multiple solutions from different threads} (2), \textit{duplicate content} (1), and \textit{archived threads} (1). 

Our inter-rater agreement for this coding task was 0.76.

While 11 participants have no problems navigating Stack Overflow, more participants complain about the amount of information they need to navigate (e.g., ``\textit{It's often finding the useful info. There are some threads where there's so much useless stuff}''). This further motivates our research for helping developers navigate Stack Overflow. 
The other challenges mentioned are also quite interesting.
For example, the difficulty in adopting solutions to a user's specific context or the need to combine multiple solutions from different threads (e.g., ``\textit{I usually have to piece together multiple stack overflow threads to find a solution to a particular situation}'').
Highlighting the context under which a solution is relevant may help alleviate the former challenge.  

\begin{findingenv}{RQ1}{rq1}
Challenges in SO navigation include sifting through lots of information, adapting solutions to the user's context, and combining solutions from multiple threads.
\end{findingenv}

\begin{table*}
\caption{Number of sentences identified by each technique, and percentage of those sentences that were highly rated}
\label{tab:selectedsentences}
\centering
\resizebox{0.85\textwidth}{!}{
\begin{tabular}{l|c|c|c|c}
\hline
\textbf{Technique}&\textbf{Number of Sentences}&\textbf{~\ref{q:q8} Highly Rated}&\textbf{~\ref{q:q9} Highly Rated}&\textbf{~\ref{q:q10} Highly Rated}\\
\hline
\hline

\texttt{\simpleif} &
\pgfkeysvalueof{number_simpleif_sentences}
&
\pgfkeysvalueof{total_pos_tech1_q8_threshold_3} 
(\printpercent[0]{\pgfkeysvalueof{total_pos_tech1_q8_threshold_3}}{\pgfkeysvalueof{number_simpleif_sentences}})
&
\pgfkeysvalueof{total_pos_tech1_q9_threshold_4} 
(\printpercent[0]{\pgfkeysvalueof{total_pos_tech1_q9_threshold_4}}{\pgfkeysvalueof{number_simpleif_sentences}})
&
\pgfkeysvalueof{total_pos_tech1_q10_threshold_4} 
(\printpercent[0]{\pgfkeysvalueof{total_pos_tech1_q10_threshold_4}}{\pgfkeysvalueof{number_simpleif_sentences}})\\
\hline

\texttt{\condinsight}& \pgfkeysvalueof{number_condinsight_sentences}
&
{\pgfkeysvalueof{total_pos_tech2_q8_threshold_3}}
(\printpercent[0]{\pgfkeysvalueof{total_pos_tech2_q8_threshold_3}}{\pgfkeysvalueof{number_condinsight_sentences}})
&
\pgfkeysvalueof{total_pos_tech2_q9_threshold_4} 
(\printpercent[0]{\pgfkeysvalueof{total_pos_tech2_q9_threshold_4}}{\pgfkeysvalueof{number_condinsight_sentences}})
&
\pgfkeysvalueof{total_pos_tech2_q10_threshold_4} 
(\printpercent[0]{\pgfkeysvalueof{total_pos_tech2_q10_threshold_4}}{\pgfkeysvalueof{number_condinsight_sentences}})\\
\hline

\texttt{\wordpattern}&\pgfkeysvalueof{number_wordpattern_sentences}
&
{\pgfkeysvalueof{total_pos_tech3_q8_threshold_3} 
(\printpercent[0]{\pgfkeysvalueof{total_pos_tech3_q8_threshold_3}}{\pgfkeysvalueof{number_wordpattern_sentences}})}
&
\cellcolor{gray!25}{\pgfkeysvalueof{total_pos_tech3_q9_threshold_4} 
(\printpercent[0]{\pgfkeysvalueof{total_pos_tech3_q9_threshold_4}}{\pgfkeysvalueof{number_wordpattern_sentences}})}
&
\pgfkeysvalueof{total_pos_tech3_q10_threshold_4} 
(\printpercent[0]{\pgfkeysvalueof{total_pos_tech3_q10_threshold_4}}{\pgfkeysvalueof{number_wordpattern_sentences}})\\
\hline

\texttt{\lexrank}&\pgfkeysvalueof{number_lexrank_sentences}
&
\cellcolor{gray!25}{\pgfkeysvalueof{total_pos_tech6_q8_threshold_3} 
(\printpercent[0]{\pgfkeysvalueof{total_pos_tech6_q8_threshold_3}}{\pgfkeysvalueof{number_lexrank_sentences}})}
&
\pgfkeysvalueof{total_pos_tech6_q9_threshold_4} 
(\printpercent[0]{\pgfkeysvalueof{total_pos_tech6_q9_threshold_4}}{\pgfkeysvalueof{number_lexrank_sentences}})
&
\cellcolor{gray!25}{\pgfkeysvalueof{total_pos_tech6_q10_threshold_4} 
(\printpercent[0]{\pgfkeysvalueof{total_pos_tech6_q10_threshold_4}}{\pgfkeysvalueof{number_lexrank_sentences}})}\\
\hline
\hline
Total Unique Sentences&
\pgfkeysvalueof{total_sentence_eval}
&
\pgfkeysvalueof{ground_truth_size_q8_limit_3} (\printpercent[0]{\pgfkeysvalueof{ground_truth_size_q8_limit_3}}{\pgfkeysvalueof{total_sentence_eval}})
&
\pgfkeysvalueof{ground_truth_size_q9_limit_4} (\printpercent[0]{\pgfkeysvalueof{ground_truth_size_q9_limit_4}}{\pgfkeysvalueof{total_sentence_eval}})
&
\pgfkeysvalueof{ground_truth_size_q10_limit_4} (\printpercent[0]{\pgfkeysvalueof{ground_truth_size_q10_limit_4}}{\pgfkeysvalueof{total_sentence_eval}})
\\
\hline
\end{tabular}
}
\vspace{-0.5cm}
\end{table*}

\section{RQ2: Potential of Highlighting Information}

\begin{figure}[t!]%
\centering
	\begin{subfigure}[t]{0.5\textwidth}%
		\centering
		\includegraphics[width=\textwidth]{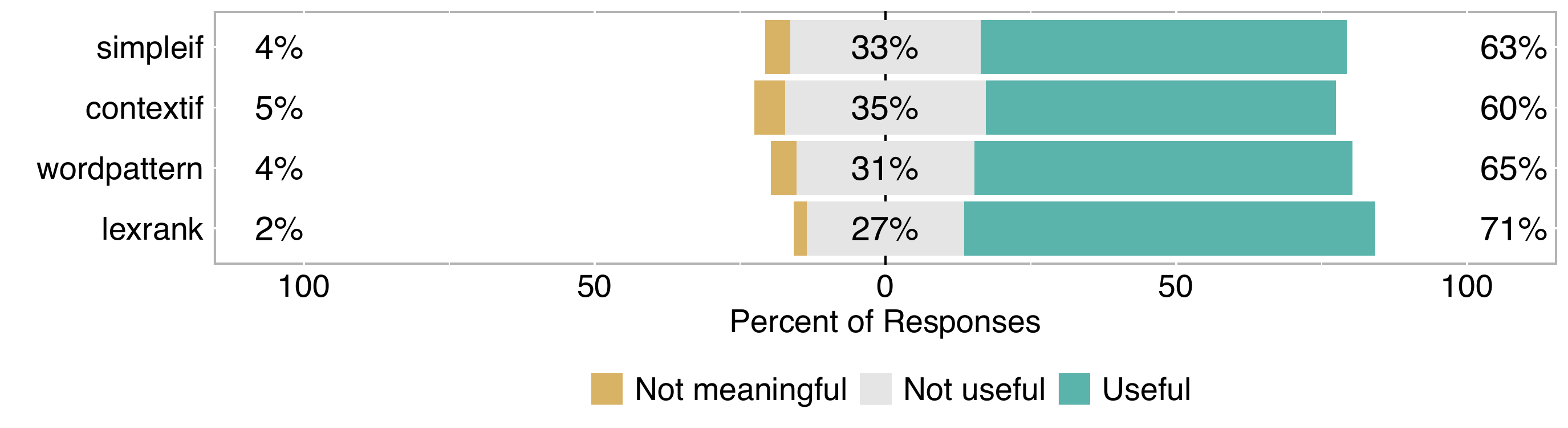}
		\caption{\ref{q:q8}}%
		\label{fig:q8-likert}	
	\end{subfigure}
	\begin{subfigure}[t]{0.5\textwidth}%
		\centering
		\includegraphics[width=\textwidth]{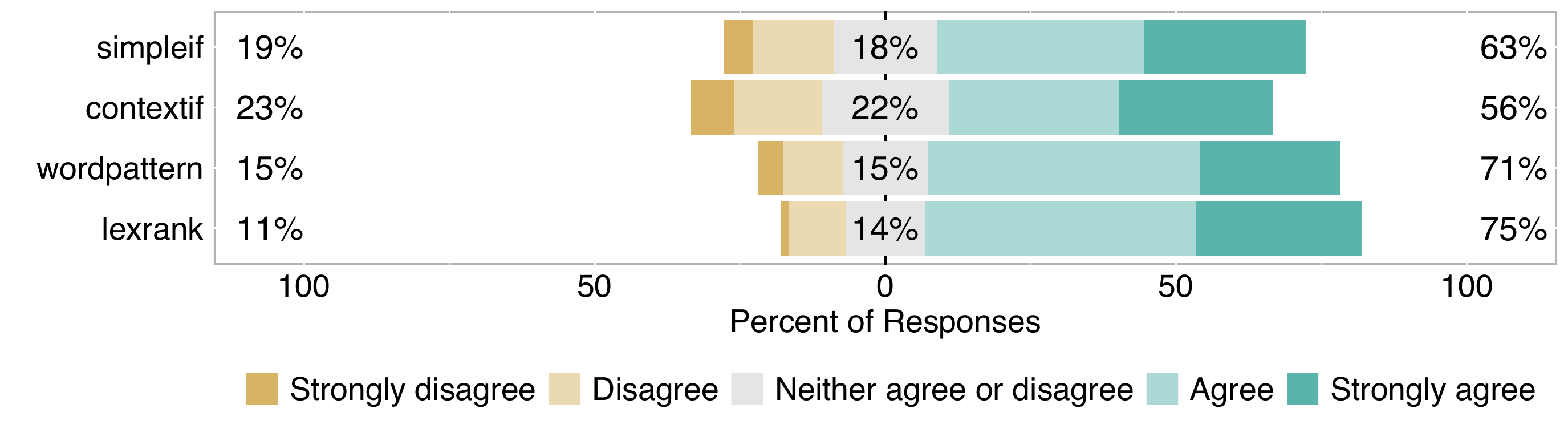}
		\caption{\ref{q:q9}}%
		\label{fig:q9-likert}	
	\end{subfigure}
	\begin{subfigure}[t]{0.5\textwidth}%
		\centering
		\includegraphics[width=\textwidth]{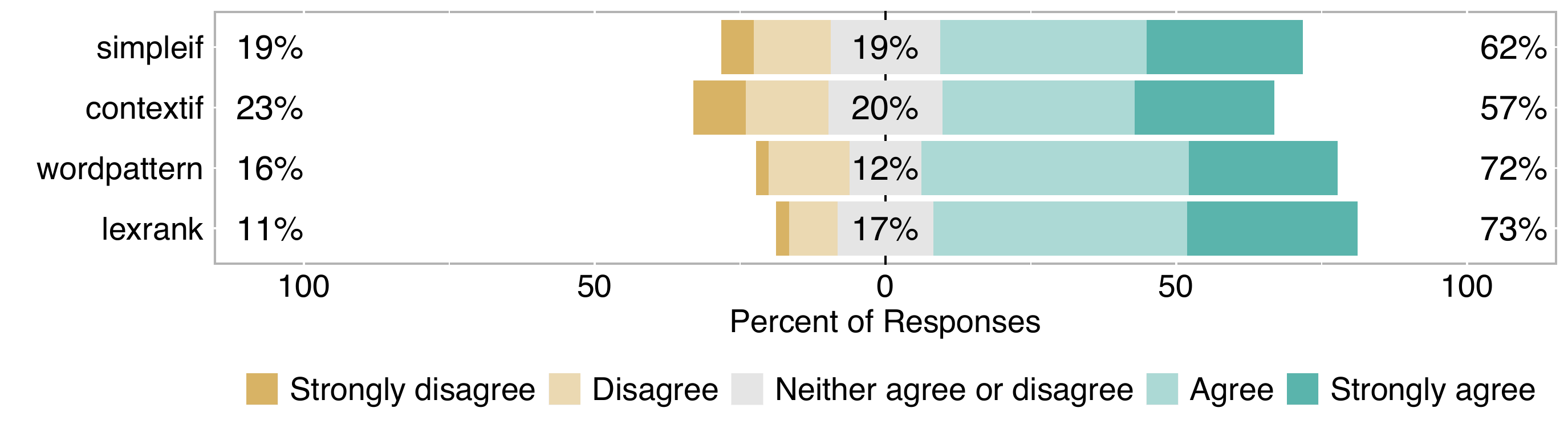}
		\caption{\ref{q:q10}}%
		\label{fig:q10-likert}	
	\end{subfigure}

	\caption{Distribution of responses per technique per question}
		\label{fig:likert}
		\vspace{-0.5cm}
\end{figure}

In RQ2, we explore the potential of highlighting information on SO. To do so, we analyze the responses to two questions: \ref{q:q15} where we explicitly asked participants what information they want highlighted and~\ref{q:q16} which explicitly asks participants whether they think highlighting conditional information could be helpful. Note that we put~\ref{q:q16} as the last question of the survey to avoid any bias. 

The codes we use for the responses in~\ref{q:q15} are:
\textit{most relevant solution} (12), \textit{direct answer} (10), \textit{relevant explanations} (8), \textit{tips} (4), \textit{code} (4), \textit{confirmed information} (2), \textit{step-by-step solution} (2), \textit{no highlighting} (2), and \textit{summary} (1).
Our inter-rater agreement for this coding task was 0.85.
Naturally, most participants wanted the most-relevant solution (e.g., ``\textit{I would like the most relevant solution to be highlighted}'') or a direct answer (e.g., ``\textit{Any links to programs or interfaces that are the direct solution}'') to be highlighted.
It is interesting to note here that the most-relevant solution to the question in the thread may not necessarily be the most relevant one to the user.
Thus, the term ``most relevant'' is relative to the specific problem the user is looking to solve, which may have slight differences to the one in the thread.
The third code in our list shows that not only are direct solutions relevant, the explanations that are needed to understand them are very important to participants. 
Previous work investigating code examples also found similar results about the importance of explanations~\cite{Nasehi2012}.

For~\ref{q:q16}, twenty participants said yes to highlighting conditional information; 11 said maybe.
The following are the codes for participant justifications (if any): \textit{yes, good to know cases} (9), \textit{yes, provides easy/quick navigation} (6), \textit{yes, depends on stack} (4), \textit{yes, with granularity selection} (1), and \textit{no, may miss big picture} (1).
Our inter-rater agreement was 0.82.
These comments suggest that the majority of participants think that highlighting conditional sentences is a good idea (e.g., ``\textit{Knowing a solution is specific to a technology, etc. will let me quickly decide if I should continue reading or look elsewhere}'').
Participants say that it is good to know the different cases when finding a solution, that it makes navigation easier, and that such highlighting is useful if the user's technology stack matches the highlighted information.
Such comments match our intuition and motivation for this work.
One participant had an interesting suggestion of allowing users to select the granularity level of the highlighting they want (e.g., programming languages, operating systems etc.): ``\textit{It might be nice to have some granularity to the highlighting that a user could specify (i.e., Do you want to highlight x, y, z?)}''.

\begin{findingenv}{RQ2}{rq2}
Participants would find highlighting the most relevant solution useful. Highlighting conditional information may be useful, especially to understand the various cases for a solution.
\end{findingenv}

\section{RQ3: Technique Comparison}
\label{sec:rq1-res}

To compare the performance of the four techniques for RQ3, we perform a quantitative analysis of the ratings from Questions~\ref{q:q8},~\ref{q:q9}, and~\ref{q:q10}.
Figure~\ref{fig:likert} shows the distribution of responses for the sentences identified by each technique for each question.
We use a two-sided unpaired Wilcoxon signed-rank test to compare the rating distribution of the various techniques and calculate effect size using $r = Z/sqrt(N)$~\cite{field2012discovering}. We use a Benjamini \& Hochberg (BH) p-value adjustment measure to account for multiple comparisons, and use $\alpha = 0.05$.
Table~\ref{tab:selectedsentences} also shows the percentage of the identified sentences that were highly rated by participants.

\paragraph*{General Quality of Sentences} In \ref{q:q8}, we asked participants to select a statement that best describes the highlighted sentence. 
While we found no statistically significant differences between the rating distributions of all four techniques, Table~\ref{tab:selectedsentences} shows that \simpleif had the highest number of highly rated sentences (32).
This is expected given that \simpleif naturally identifies more sentences given its simple criteria.
However, a higher percentage of the sentences identified by \lexrank end up being highly rated by participants, in this case seen as meaningful and useful. 

\paragraph*{Desire to Locate Sentences}
In~\ref{q:q9}, we ask participants if they would like to quickly locate the highlighted sentence when navigating the thread.
Again, we find no statistically significant differences between the distribution of ratings across the techniques.
However, as shown in Table~\ref{tab:selectedsentences}, the number of highly rated sentences in each technique differs between~\ref{q:q8} and~\ref{q:q9}. 
We point out that the highly rated sentences in~\ref{q:q9} are not necessarily a subset of those in~\ref{q:q8} due to the nature of the question.
Question~\ref{q:q8} asked participants to rate the sentence on its own rather than for a specific purpose/goal, while \ref{q:q9} specifically asks about the goal of quickly locating a given sentence.
One example of a sentence that was positively rated in~\ref{q:q8} but negatively rated in~\ref{q:q9} is \textit{``And if so it returns the default value''} from answer \post{50523464}. 
This sentence is part of a paragraph that explains the logic of a piece of code that preceded it.
While the sentence is meaningful and may be helpful as an explanation of the problem and provided code, someone navigating this thread would not be keen about quickly locating this sentence per se since it does not give any information about the context/topic/value of the answer.
On the other hand, an example of a sentence that was less positively rated in~\ref{q:q8} but positively rated in~\ref{q:q9} is \textit{``You're not looping over an Array, you are looping over the properties of an Object with a \href{https://developer.mozilla.org/en-US/docs/Web/JavaScript/Reference/Statements/for...in}{for...in}  loop.'} from thread \post{50958104}. 
While this sentence does not provide a direct solution to the problem, it explains why the problem occurs.
As one participant elaborates \textit{``The highlighted sentence gives an explanation to the problem but is not suggesting any solution to the problem. But, still highlighting it might help me in navigating to that solution which might also contain the actual solution to the problem.''}

Table~\ref{tab:selectedsentences} shows that \wordpattern has the highest proportion of highly rated sentences for~\ref{q:q9}.
It is also worth noting that \condinsight has a smaller number of highly rated sentences when compared to \simpleif. 
Given that \condinsight is a subset of \simpleif, this suggests that some of our filtering criteria may have been too strict, resulting in filtering out sentences that are valued by participants.
We discuss this phenomenon in more detail in Section~\ref{sec:discussion}, as well as look more closely at the overlap between the sentences identified by the four techniques.

\paragraph*{Helpfulness In Identifying Relevant Solutions}
In \ref{q:q10}, we ask participants whether the given sentence helps them quickly identify relevant solutions and disregard irrelevant ones.
This is the most important question for our goal of providing useful \navcue{s} to Stack Overflow users.
Table~\ref{tab:selectedsentences} shows the percentage of sentences identified by each technique that were highly rated by participants.
Again, there may be sentences that are highly rated in this question but not in other ones.
For example, the following sentence \textit{``But if you have an \texttt{include} property in your \texttt{tsconfig.json}:''} from thread~\post{51494250} has a median rating $>=4$ for~\ref{q:q10} but not~\ref{q:q9}.
As one participant explains their rating for this sentence, \textit{``This explains one situation in which this person's solution might be helpful''}.
Thus, when specifically asked if this sentence would help users quickly identify relevant solutions and disregard irrelevant ones, more participants rated this sentence more positively.

Following this logic, as shown in Table~\ref{tab:selectedsentences}, we can see that the number of \condinsight sentences that were positively rated increased between~\ref{q:q9} and~\ref{q:q10}, suggesting that these conditional sentences do indeed serve the purpose of differentiating between solutions better.
However, \lexrank still outperforms the other techniques in terms of having the highest ratio of its sentences rated highly.
Note that we find a statistically significant difference between the rating distribution of \condinsight and \lexrank ($p=0.045$), but with a very small~\cite{sawilowsky2009new} effect size of 0.155. 

\begin{findingenv}{RQ3}{rq3}
When considering a median rating of at least ``Agree'', a higher percentage of \lexrank's sentences are perceived as helpful navigational cues.
\end{findingenv}

\section{RQ4: Helpful Navigation Information}
\label{sec:rq2-res}

We now present the results of our qualitative analysis to answer RQ4. 
We present the details of the final set of codes used, the inter-rater agreement for each question, and the number of instances per code.
For all questions, we find that the majority of instances were assigned exactly one code, with a max number of codes of 3.

\paragraph*{Sentence Analysis} To understand the information participants look for in navigation cues, we look at the \pgfkeysvalueof{ground_truth_size_q10_limit_4} unique highly rated sentences for Question~\ref{q:q10}.
Our goal is to identify the kind of information in the sentence, or its role in the thread.
Thus, we also use participants' rating justifications from~\ref{q:q11} to guide us during the process.

The set of codes we use, along with the number of instances per code in parentheses is as follows:  \textit{explanation} (23), \textit{specific condition} (19), \textit{API note} (17), \textit{direct solution} (5),  \textit{label} (5), and \textit{other} (1). 
Our inter-rater agreement kappa score is 0.78.
Looking at the top three codes, an example of an explanation sentence is \textit{``If it returned a falsy value (i.e. 0) then the value of the other operand of $||$ will be returned (which is sorting by timestamp).''}
Here, the poster is explaining some concept related to the question, or in other cases related to the provided solution.
For explanation sentences, participants use justifications that involve the words ``explain'', e.g., \textit{``That's the simplest and clear[est] explanation.''}
An example of a sentence highlighting a specific condition is \textit{``You can use session storage if you want the data to be retrieved once PER SESSION or local storage if you want to have better control of the data's `expiration'.''}
Participants provided justifications for this category of sentences such as \textit{``This single sentence answers the question thoroughly, with two separate options.''}
Sentences containing API notes typically provided extra information about a specific method.
An example is \textit{``Another thing to notice, is that \texttt{require} is synchronous, so if your JSON is specially large, the first time you instantiate \texttt{MyClass} the event loop will be blocked.''}
One of the participants comments on this sentence saying, \textit{``This type of information would allow me to avoid a future bug or unexpected behavior, and directly relates to the question.''}

\paragraph*{General reasons for high ratings} In~\ref{q:q12}, we asked participants about what generally made them rate a sentence highly.
Recall that participants answer this question at the end of the survey, allowing them to reflect on all sentences they evaluated.
The set of codes we use, along with the number of instances per code in parentheses is as follows:  \textit{direct solution} (16), \textit{explanation} (13), \textit{relevance of info} (11), \textit{code/lib} (6), \textit{well-written} (5), \textit{SO info} (2), \textit{alternate solution} (1), and \textit{warning} (1).
Our inter-rater agreement kappa score is 0.86.

Our codes show that participants explicitly mention the existence of a direct solution or explanation the most. The quality of the sentence itself, in terms of how well-written or concise it is, also affected their rating.
Several participants find that a sentence is useful when it mentions a specific library, or piece of code (e.g., an API).
It is interesting to see that only one participant explicitly mentions alternatives, while based on our previous analysis of what information is contained in highly rated sentences, 19 out of the \pgfkeysvalueof{ground_truth_size_q10_limit_4} analyzed sentences contained some form of a specific condition.
We speculate that participants do not necessarily refer to these as conditional sentences, or alternative solutions, but instead find them useful because they mention particular contexts, whether programming languages, library names, or APIs.

\paragraph*{General reasons for low ratings} We also look at the answers to~\ref{q:q13}, which asked about general reasons for low ratings.
The set of codes we use, along with the number of instances per code in parentheses is as follows: \textit{not relevant to main issue} (23), \textit{filler sentences} (7), \textit{no meaningful information} (6), \textit{speculation} (4), \textit{not standalone} (2),  \textit{formatting} (1), \textit{complicated} (1), \textit{contains no code} (1), \textit{generic} (1), and \textit{other} (1).
Our inter-rater agreement kappa score is 0.87.

As can be seen, sentences that did not contain information relevant to the main issue being asked in the thread, such as direct solutions, were seen as not useful.
Filler sentences are farewell sentences or just social comments, such as people saying good bye or good luck at the end of a post.
Sentences with no meaningful information of how to apply the given information to solve the problem were also rated low.

\begin{findingenv}{RQ4}{rq4}
The majority of highly rated sentences in~\ref{q:q10} contain explanations or specific conditions. The most mentioned factor affecting participants' sentence ratings is whether they directly solve the main issue in question.
\end{findingenv}

\section{Discussion}
\label{sec:discussion}

\begin{figure}[t!]%
\centering
\includegraphics[width=0.3\textwidth]{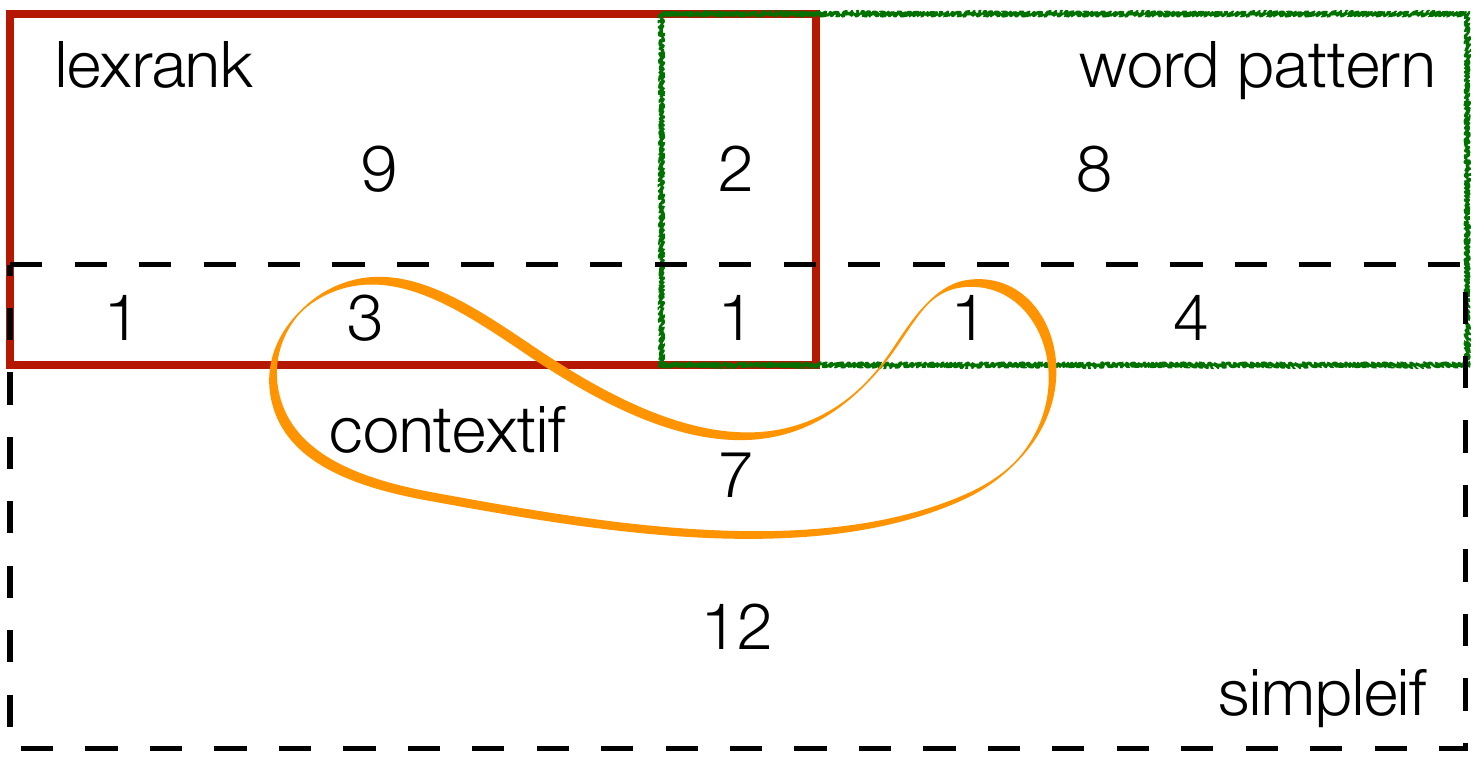}
\caption{Venn diagram illustrating the techniques corresponding to the \pgfkeysvalueof{ground_truth_size_q10_limit_4} highly rated sentences in Question~\ref{q:q10}}%
\vspace{-0.6cm}
\label{fig:q10_highlyratedvenn}	
\end{figure}

Our goal in this work is to investigate what kind of information is helpful for developers to navigate through potential solutions on Stack Overflow, and possible techniques to identify this information.
Our work is the first to investigate essential sentences for the purpose of navigating technical information on Stack Overflow.
Our qualitative results suggest that we are on the right track in thinking that contextual information may be useful.
We believe that techniques that identify and highlight such contextual information can provide valuable navigation cues on Stack Overflow.
Identifying conditions and contexts related to a query can also help users contextualize their search and reach their target thread more easily.
For example, if we identify that there is a big variation in answers to a given task based on the programming language or operating system, we can prompt the user to select one of the existing contexts during their search.
Finally, this information can also help with automated thread or answer tagging on Stack Overflow, which would make both manual navigation as well as automated search engines more effective.

Instead of re-inventing the wheel, we looked for existing techniques that may potentially find relevant information for our purposes, namely \lexrank and \wordpattern.
Our results show that both techniques are indeed promising, since a high proportion of the sentences they identify is highly rated by participants.
Additionally, given the idea of contextual information, we focused on conditional sentences since conditions may likely signal specific contexts.
In the \condinsight technique, we designed some heuristics to identify meaningful conditions.
However, the results show that our heuristics were rather strict, where they filtered out many \simpleif sentences that ended up being highly rated by participants.
We plan to investigate if ignoring parts of speech in the condition helps and if other means for differentiating technology related conditions can be used~\cite{yan2019leveraging,chan2003dynamic, Nassif:2019}.

Not all sentences that help users navigate a thread are necessarily providing context information. 
For example, the following sentence, detected by \lexrank, ``\textit{You are trying to cast to \texttt{[String: Any]}, but you have an array of \texttt{[String: Any]} because your response enclosed in \texttt{[]} brackets.''} was highly rated. Since it does not contain any conditional phrases, it was not picked up by \simpleif or \condinsight. However, participants found it relevant since it provided a direct explanation of the problem being faced in this thread. We conclude that there may be different types of navigation issues at hand. The first is a user who knows the particular thread that describes their exact problem and just wants to quickly find (and understand) the direct solution. In this case, a sentence such as the above one is what they need. The second is a user who is facing a similar problem but in a very specific context and is browsing threads to try to find someone who shares this context. In this case, sentences that contextualize the solution (and thus likely contain some form of condition) are more useful. Our survey did not differentiate these types of users since we did not provide them a concrete task to do. 
Thus, we conclude that both types of information are useful, but may depend on the use case. 

Figure~\ref{fig:q10_highlyratedvenn} illustrates this point further.
We show all the \pgfkeysvalueof{ground_truth_size_q10_limit_4} sentences that were highly rated in Question~\ref{q:q10}, and the corresponding techniques that identified them, including any overlap.
As can be seen,  the majority (\pgfkeysvalueof{total_pos_tech1_q10_threshold_4}/\pgfkeysvalueof{ground_truth_size_q10_limit_4} = 60\%) of these highly rated sentences are conditional sentences (i.e., \simpleif sentences), while the remaining sentences are identified by \wordpattern or \lexrank. 
Additionally, with the exception of a very small effect size for the difference between \lexrank and \condinsight, we found no statistically significant difference between the four techniques.
Thus, we can conclude that there is no single technique that is clearly a ``winner'', and a future direction could be investigating meaningful combinations as well as additional techniques.

Finally, it is worth noting that the highly rated sentences did not necessarily exist in the accepted answer for the thread or in the highest scoring answers. 
For example, the sentence \textit{``I'd only put \texttt{require} inside the constructor if it was a dynamic dependency, [...]''} had a median rating of  5, even though the corresponding answer's score was 1.
This supports our belief that developers often look for information relevant to their context, which may exist in various answers in a thread.

\section{Threats to Validity}

\paragraph*{Internal Validity}
There may be additional essential sentences useful for \navcue{s} that none of the techniques detected.
Thus, our comparison of techniques is a relative one with respect to the four examined techniques.

The choice of the threshold used for  the definition of ``highly rated'' affects the results we obtain.
We chose a threshold that indicates a high rating, based on the meaning of each question.

Participants may provide random answers to the survey questions, including the quality gate question in which it will be correct in 40\% of the time. This is always a threat to any survey, and we tried to mitigate it as much as possible.

We rely on the Stanford CoreNLP toolkit for the processing of sentences.
Any inaccuracies in it may lead to inaccuracies in the presented techniques.

When sampling from Stack Overflow, we did not apply any filtering based on answer score, because an essential sentence does not necessarily exist only in highly scored answers. As discussed in Section~\ref{sec:discussion}, we found highly rated sentences in answers with score 1. Additionally, we found no Spearman rank correlation between any of the ratings and answer score.

\paragraph*{Construct Validity} The two existing techniques we used, \wordpattern and \lexrank, were not designed for the purpose of SO navigation cues.
Our goal is to investigate how applicable existing techniques are to identifying \navcue{s} and to identify future steps.

Our survey participants were not provided with a concrete task to think of when evaluating threads. This was intentional since creating a concrete task with a specific context that does not bias the results towards the conditional sentences in the thread is difficult.
Without a concrete task/context, a controlled experiment setting to compare the highlighting of essential sentences to current navigation techniques is infeasible.
Participants might rate the sentences differently if they were provided with a specific context.
Our survey mitigates that by asking about different aspects of the sentences, as well as eliciting free-form feedback that we qualitatively analyzed.

To ensure that we can statistically compare the techniques and to avoid burdening participants, we evaluated threads with a similar number of extracted sentences by each technique.
This sampling may mean that we evaluated on a unique subset of threads since the majority of threads had sentences identified by only some of the techniques.

\paragraph*{External Validity}
There are other language patterns that encode conditional information, e.g., ``For Linux, use ...'' Thus, the current \simpleif and \condinsight sentences do not necessarily represent all contextual information. To investigate the idea of using conditional information for \navcue{s}, we started with the simple form of sentences with ``if''.

Our results are based on the evaluation of \pgfkeysvalueof{total_threads_eval} json threads, with \pgfkeysvalueof{total_sentence_eval} highlighted sentences, by \pgfkeysvalueof{final_num_participants} participants, and may not generalize beyond that.
That said, our survey participants have diverse background, with varying expertise levels with json as well as varying occupations and programming experience, which gives us confidence that our results are not biased towards the views of a particular population sample.

\section{Conclusion}

Given the amount of information available on Stack Overflow, identifying which part of which answer is suitable for a user's task and context is difficult.
In this paper, we compared four techniques to identify essential sentences providing navigational cues for guiding users to the most suitable (part of an) answer. 
Our results show that a high percentage of the sentences identified by a text summarization technique, \lexrank, are highly rated.
However, we also show that the other techniques find many additional highly rated sentences, which suggests that there is no clear silver bullet.
We find that most of the highly rated sentences contain explanations or specific conditions.
The results support our intuition that conditional or contextual information can help users quickly navigate information on Stack Overflow.
We plan to investigate additional heuristics to further differentiate between useful contextual information and irrelevant conditions.
We publicly share all the code and data from our work~\cite{artifact}, which also includes the collected sentence ratings from our survey evaluation that can be used to guide future work.

\section*{Acknowledgements}
Thanks to Benyamin Noori for early investigation of conditional sentences and to Samer Al Masri for implementing the survey website. This research was undertaken, in part, thanks to funding from the Canada Research Chairs program and the Australian Research Council’s Discovery Early Career Researcher Award (DECRA) funding scheme (DE180100153). This work was inspired by the International Workshop series on Dynamic Software Documentation, held at McGill’s Bellairs Research Institute.

\balance

\end{document}